\documentclass[review]{elsarticle}
\usepackage{mdframed}
\usepackage{amsmath,amssymb}
\usepackage{lipsum}
\usepackage{bm}
\usepackage{graphicx}
\usepackage{graphics}
\usepackage{amsmath}
\usepackage{array}
\usepackage[caption=false]{subfig}
\usepackage{xcolor}
\usepackage{subfig}
\usepackage{hyperref}
\usepackage[capitalise]{cleveref}
\usepackage{textcomp}
\usepackage{epstopdf}
\usepackage{cases}
\usepackage{amssymb}
\usepackage{multirow}
\usepackage{siunitx}
\usepackage{cases}
\usepackage{tabularx}
\usepackage[top=1.1in, bottom=1.1in, left=1.0in, right=1.0in]{geometry}
\usepackage[utf8]{inputenc}
\usepackage{tikz}
\usepackage{comment}
\usepackage{lineno,hyperref}
\modulolinenumbers[5]
\usepackage{booktabs}
\begin{document}

\title{Kinetic simulation of magnetic-field-tuned hydrodynamic electron transport in graphene corbino disk}
\author[add1,add2]{Chuang Zhang}
\ead{zhangc26@zju.edu.cn}
\author[add3]{Meng Lian}
\ead{lianmeng@hust.edu.cn}
\author[add4]{Hong Liang}
\ead{lianghongstefanie@163.com}
\author[add5]{Xiaokang Li}
\ead{lixiaokang@hust.edu.cn}
\author[add6]{Zhaoli Guo\corref{cor1}}
\ead{zlguo@hust.edu.cn}
\author[add3]{Jing-Tao L\"u\corref{cor1}}
\ead{jtlu@hust.edu.cn}
\address[add1]{Center for Digital-Physics Interdisciplinary Research, Zhejiang University, Hangzhou 310058, China}
\address[add2]{Institute of Fundamental and Transdisciplinary Research, Zhejiang University, Hangzhou 310058, China}
\address[add3]{School of Physics, Institute for Quantum Science and Engineering and Wuhan National High Magnetic Field Center, Huazhong University of Science and Technology, Wuhan 430074, China}
\address[add4]{Department of Physics, School of Sciences, Hangzhou Dianzi University, Hangzhou 310018, China}
\address[add5]{Wuhan National High Magnetic Field Center and School of Physics, Huazhong University of Science and Technology, Wuhan 430074, China}
\address[add6]{Institute of Interdisciplinary Research for Mathematics and Applied Science, Huazhong University of Science and Technology, Wuhan 430074, China}
\cortext[cor1]{Corresponding author}

\date{\today}
\begin{abstract}

{\color{blue}{Hydrodynamic electron transport, in which electrical transport in solids resembles fluid hydrodynamics when momentum-conserving electron-electron scattering dominates, has attracted much attention over the past decade.
However, its thermal aspects have received considerably less attention.
In this paper, electron transport in a graphene Corbino disk is systematically simulated by solving the stationary Boltzmann transport equation with a dual-relaxation-time Callaway model, where momentum-conserving and momentum-relaxing scatterings are explicitly distinguished. 
By varying the magnetic field intensity and the scattering rates, the electric charge and heat flux responses are compared across the diffusive-to-hydrodynamic crossover under electric-field or temperature-gradient drives. 
It is shown that magnetic-field-induced deflection of both fluxes is strongly enhanced in the hydrodynamic regime but nearly suppressed in the diffusive regime. 
Under electric-field driving, a pronounced temperature rise is observed in the hydrodynamic regime due to reduced dissipation, while the diffusive regime remains nearly isothermal. 
Under temperature-gradient driving, the deflection is reversed relative to the electric-field case. 
These findings establish that thermal behaviors could provide a sensitive and independent diagnostic of electron hydrodynamics, with the magnetic field being identified as an effective discriminator between collective and dissipative conduction.}}

\end{abstract}
\maketitle

\section{Introduction}

Hydrodynamics phenomena originate from the strong interactions of microscopic (quasi)particles, where the particle number, (quasi)momentum and energy are conserved and a local equilibrium exists during the scattering process~\cite{hydrodynamic_2023charge_review,hydrodynamic_electron_review,para_hydrodynamics_flakes_2023,beardo2026phonon}.
For example, the fluid hydrodynamic phenomena are ubiquitous in daily life, such as turbulence, vortices, Poiseuille flow, wave and so on.
These universal hydrodynamic behaviors do not depend on the types of quasiparticles and their microscopic interaction~\cite{hydrodynamic_electron_review,eDUGKS_2024,zhu2025out}. 
Theoretically, they can be described by mesoscopic Boltzmann transport equation (BTE)~\cite{chandra2019a,hydrodynamic_electron_review,pnas_superballistic2017,sulpizio2019visualizing} or macroscopic hydrodynamic equations in the continuum limit~\cite{PhysRevLett.130.166201,PhysRevB.105.155307,PhysRevLett.123.026801}.

However, hydrodynamic behaviors of electrons in crystal solids at room temperature are not easy to be observed because the crystal momentum is usually not conserved during the electron scattering processes~\cite{hydrodynamic_2023charge_review,hydrodynamic_electron_review}. 
In $1960$s, Gurzhi~\cite{Gurzhi_1963,Gurzhi_1968} theoretically proposed that, as a signature of electron hydrodynamics, electrical resistance of a conductor within certain size and temperature ranges could decrease when the temperature increases. 
The electron hydrodynamic phenomena can be predicted in a certain low temperature window when the momentum-conserving (MC) electron-electron scattering process is much sufficient than other momentum-relaxing (MR) scattering process including scattering with impurities, phonons and so on.
This pioneering work opened the door of electron hydrodynamics in solid materials, which was expected to significantly modify or improve the electric and thermal transport coefficients~\cite{PhysRevB.21.3279}. 
Although electron hydrodynamics has been studied theoretically over half a century~\cite{PhysRevB.51.13389,Gennaro_1984,PhysRevLett.74.3872}, only a little experimental progress has been made in the $20$th century~\cite{PhysRevLett.52.368,PhysRevB.51.13389}.

In the past decade, the study of electron hydrodynamics enjoyed a renaissance~\cite{ella2019,hydrodynamic_2023charge_review,hydrodynamic_electron_review,PhysRevLett.113.235901} due to the discovery of new materials, e.g., graphene~\cite{PhysRevB.92.115426,pnas_superballistic2017,PhysRevLett.122.137701,PhysRevLett.121.236602}, in which the electron-phonon coupling is weak and MC scattering process is much stronger.
Many electron hydrodynamic phenomena have been measured~\cite{nature_2022_hydrodynamics,vool2021imaging,jaoui2021thermal,viscous_2023_hydrodynamics,ku2020imaging,PhysRevLett.129.157701,bandurin2018}, such as the electron Poiseuille flow~\cite{sulpizio2019visualizing}, negative nonlocal resistance or vortices/whirlpool~\cite{whirlpool_2024_science,bandurin2016b,PhysRevB.97.245308,science_2016_hydrodynamicsPdCo,aharon2022direct,stokes_hydrodynamic_2019}, Hall viscosity~\cite{berdyugin2019a}, violation of Wiedemann-Franz law~\cite{science_2016_breakWFlaw,PhysRevLett.130.166201}, super-ballistic flow~\cite{krishnakumar2017} and so on.
The difference of macroscopic behaviors between hydrodynamic, ballistic and diffusive electron transport are widely studied in various materials or geometries, both theoretically~\cite{estrada2024alternative,scaffidi2017,levitov_electron_2016,chandra2019a,PhysRevB.95.115425} and  experimentally~\cite{whirlpool_2024_science,sulpizio2019visualizing,bandurin2018,krishnakumar2017,PhysRevB.97.245308,science_2016_hydrodynamicsPdCo,aharon2022direct}. 
Levitov and Falkovich used the linearized electronic Navier-Stokes equation to study the current vortices and viscosity~\cite{levitov_electron_2016}.
Results show that vortices could appear in the hydrodynamic regime while disappear in the diffusive regime.
Similar current vortices, viscosity or whirlpool phenomena have also been predicted in various rectangle or Corbino disk geometries by applying voltage in different positions to control current flow~\cite{PhysRevResearch.7.L022029,PhysRevResearch.7.013087,corbino2024,PhysRevB.107.235401,aharon2022direct,anisotropic_hydrodynamics_2020,anomalous_2023_corbino,stokes_hydrodynamic_2019,falkovich_linking_2017,torre2015,chandra2019a,PhysRevB.94.155414,PhysRevB.97.245308,PhysRevX.11.031030,PhysRevB.105.155307}. 
{\color{blue}{For instance, Shavit, $et~al$.~\cite{PhysRevLett.123.026801} reported that the nonlocal relation between the current and electric field due to momentum-conserving interparticle collisions leads to a total or partial field expulsion from such flows, which results in freely flowing currents in the bulk and a boundary jump in the electric potential at current-injecting electrodes. 
Li, $et~al$. obtained thermoelectric coefficients of the system in the crossover region between charge neutrality and high electron density regime~\cite{PhysRevB.105.125302}. 
The thermal conductance exhibits a sensitive Lorentzian dependence on the electron density.
Gall, $et~al$. found that local temperature and electric potential are discontinuous at the interfaces with the leads as well as the device resistance in neutral graphene~\cite{PhysRevB.107.045413}.
They also reported the results of a comprehensive study of the interplay of viscosity, disorder-induced scattering, recombination, energy relaxation, and interface-induced dissipation based on the fully consistent hydrodynamic description~\cite{PhysRevB.107.235401}. 

Macroscopic models have been widely used in above studies and made success in capturing the qualitative and in many cases, quantitative features of the electron hydrodynamics.
However, these macroscopic approaches rely on continuum or local equilibrium assumption as well as classical constitutive relations, such as the Navier–Stokes or Ohm–Stokes equations.
They cannot access the full momentum-resolved information encoded in the electron distribution function, nor can they seamlessly interpolate between the diffusive and hydrodynamic regimes without invoking additional assumptions~\cite{yao2026imaging,zhangx_2026_thermoX}. 
A mesoscopic kinetic approach, based on the Boltzmann transport equation (BTE), offers a complementary and more fundamental perspective~\cite{PhysRevResearch.7.013087,chandra2019a,hydrodynamic_electron_review,pnas_superballistic2017,sulpizio2019visualizing}. 
It explicitly describes the evolution of the electron distribution function under electric and magnetic fields, temperature gradient, MR/MC scattering processes and boundary conditions, providing direct insight into the microscopic origins of macroscopic transport phenomena. 
Such an approach naturally encompasses all transport regimes within a unified numerical framework, without requiring a priori assumptions about the validity of hydrodynamic or continuous descriptions.}}
In addition, compared to electric behaviors of hydrodynamic electron transport~\cite{levitov_electron_2016,hydrodynamic_2023charge_review,hydrodynamic_electron_review,NAROZHNY2019167979}, its thermal properties got less attention~\cite{gooth2018thermal,PhysRevB.105.L241405,PhysRevB.105.125302,science_2016_breakWFlaw,PhysRevLett.130.166201}.
Actually apart from the electrical properties including mobility and conductivity~\cite{PhysRevLett.52.368,PhysRevB.51.13389}, thermal performance is also critical for the thermal management problem under the condition of ultra-high heat flux~\cite{chen_non-fourier_2021,yao2026imaging,zhangx_2026_thermoX} in the current development of semiconductor devices.
Electrical and thermal transport affect each other~\cite{PhysRevB.105.L241405,PhysRevB.105.125302,science_2016_breakWFlaw,PhysRevLett.130.166201}, therefore it is necessary to simultaneously study electrical and thermal properties of hydrodynamic electron transport.

In this study, the thermal behaviors of hydrodynamic electron transport in a homogeneous Corbino disk geometry under the magnetic field are studied.
The electron BTE with a dual-relaxation-time Callaway model~\cite{PhysRevResearch.7.013087,chandra2019a,hydrodynamic_electron_review,pnas_superballistic2017,sulpizio2019visualizing} is adopted to systematically study electron transport in a graphene Corbino disk. 
An implicit discrete ordinate method is developed to solve it numerically crossover the hydrodynamic and diffusive regimes.
The Newton method is used to solve the nonlinear relationships between the Fermi-Dirac distribution, chemical potential and temperature.
The competition effects on hydrodynamic transport among magnetic field, electric field and temperature gradient are investigated within different scattering rates.
Electric and thermal properties of both diffusive and hydrodynamic electron transport in the spatial domain are compared and discussed.
The rest of paper is organized as follow. In Sec.~\ref{sec:eBTE}, the kinetic model and associated numerical methods are introduced. Then a systematical discussion is conducted in Sec.~\ref{sec:discussions}. Finally, a conclusion is made in Sec.~\ref{sec:conclusions}.

\section{Model Equation and Method}
\label{sec:eBTE}

\subsection{Electron Boltzmann transport equation}

Electron transport at steady state can be described by the electron Boltzmann transport equation (eBTE) in the semiclassical limit~\cite{chandra2019a,hydrodynamic_electron_review,pnas_superballistic2017,sulpizio2019visualizing,eDUGKS_2024},
\begin{align}
\bm{v} \cdot \frac{\partial f }{\partial \bm{x} } + \bm{F} \cdot  \frac{\partial f }{\partial \bm{p} } &= - \frac{ f -f_{mc}^{eq} }{\tau_{mc} }  - \frac{ f -f_{mr}^{eq} }{\tau_{mr} }, 
\end{align}
which describes the evolution of electron distribution function $f$ with the spatial position $\bm{x}$ and momentum $\bm{p}= \hbar \bm{k}$ under the external force $\bm{F} =e \left( \bm{E} + \bm{v} \times \bm{B}  \right)$ of electric field $\bm{E}=- \nabla \varphi$ and magnetic field $\bm{B}$, where $e=-1.6 \times 10^{-19}$ C is the electron charge, $\hbar$ is the reduced Planck constant and $\varphi$ is the electric potential.
Electron group velocity $\bm{v} = \partial \varepsilon /\partial {\bm{p}}$ depends on the electronic energy band structure $\varepsilon=\varepsilon(\bm{p})$. 
{\color{blue}{The actual scattering process is extremely complex, involving interactions between electrons-electrons, electrons-phonons~\cite{APR2026_phonondrag} and electrons-impurities, etc. 
To simplify the calculation, the classic double relaxation time Callaway model is adopted and the phonon transport or scattering mechanisms are ignored in this paper.
The Callaway model has played a significant role in previous studies on the hydrodynamic electron transport, which splits the entire scattering mechanism into two parts. 
One part is the MC scattering process that satisfies momentum conservation, and the other part is the MR scattering process that does not satisfy momentum conservation, including defects, isotope, disorder, impurities scattering.
Both of these scattering processes satisfy the conservation principles of energy and particle number density, and are expressed by relaxation time approximation with $\tau_{mc}$ and $\tau_{mr}$, respectively.}}
The MR scattering relaxes the electron distribution to the Fermi-Dirac distribution~\cite{kaviany_2008},
\begin{align}
f_{mr}^{eq}(\varepsilon,\mu,T)=\frac{1}{ \exp{ \left( \frac{\varepsilon - \mu}{k_B T} \right) } +1  } , 
\end{align}
where $T$ is the temperature, $\mu$ is the chemical potential, $k_B$ is the Boltzmann constant.
The equilibrium state of MC scattering process is
\begin{align}
f_{mc}^{eq}(\varepsilon,\mu,\bm{u},T)=\frac{1}{\exp{\left(\frac{\varepsilon-\mu-\bm{p} \cdot \bm{u} }{k_B T} \right) } +1  },
\end{align}
where $\bm{u}$ is the macroscopic drift velocity.

{\color{blue}{In this paper, we assume that the variance of the electron number density throughout the entire system is very small, and the spatial distributions of the electric potential is almost determined by the boundary conditions and the Laplace equation $\nabla^2 \varphi \approx 0$.
We adopt a relative reference frame based on the conduction band bottom. 
In this way, the band does not change with the spatial position. 
The chemical potential $\mu$ is determined by the electron number density, while the electric potential $\varphi$ determines the spatial distributions of the electric field. 
The combination of these two can yield the electrochemical potential, which will not be extensively discussed in this article.}}

Macroscopic variables including particle number density $n$, electric current $\bm{j}$, energy density $U$ and heat flux $\bm{q}$, can be obtained by taking the momentum of distribution function in the framework of kinetic model,
\begin{align}
n = \left< f \right>,  \quad  \bm{j} = e \left< \bm{v}  f \right>,  \quad 
U = \left< \varepsilon  f \right>,  \quad \bm{q}= \left< \bm{v}  \left( \varepsilon- \mu \right)  f \right> ,
\end{align}
where $\left<\right>$ represents the integral over the whole momentum space.
The local temperature $T$ and chemical potential $\mu$ could be updated by assuming a local equivalent equilibrium,
\begin{align}
\left< f \right> = \left< f^{eq}_{mr} ( \mu ,T  )  \right>,   \quad 
\left< \varepsilon  f \right> =\left< \varepsilon f^{eq}_{mr} ( \mu ,T  ) \right>.  
\end{align}
Particle number density and energy density are conserved for both MC and MR scattering process, and momentum is also conserved for MC scattering process so that
\begin{align}
\left<  \frac{ f -f_{mr}^{eq}( \mu_{mr},T_{mr} ) }{ \tau_{mr}(T) }  \right> = \left<  \varepsilon \frac{ f -f_{mr}^{eq}( \mu_{mr},T_{mr} ) }{ \tau_{mr}(T) }  \right> = 0, \\
\left<  \frac{ f -f_{mc}^{eq} ( \mu_{mc},T_{mc},\bm{u} ) }{ \tau_{mc}(T) }   \right> =\left<   \varepsilon \frac{ f -f_{mc}^{eq} ( \mu_{mc},T_{mc},\bm{u} ) }{ \tau_{mc}(T) } \right> = \left<  \bm{p}  \frac{ f -f_{mc}^{eq}( \mu_{mc},T_{mc}, \bm{u} ) }{\tau_{mc}(T) }  \right> =0 ,
\end{align}
where $\mu_{mr}$, $T_{mr}$, $\mu_{mc}$, $T_{mc}$ are invoked to ensure the conservation principle of scattering processes~\cite{chandra2019a}.
When $\tau_{mc}$ and $\tau_{mr}$ are independent of momentum, $T_{mc}=T_{mr}=T$, $\mu_{mr}=\mu_{mc}=\mu$.
Newton method is used for the above nonlinear equations.

\subsection{Implicit discrete ordinate method}

To solve the stationary eBTE iteratively, a classical implicit discrete ordinate method is used, where the semi-implicit scheme is used for the scattering terms and fully-implicit scheme is used for the gradient of distribution function in both the spatial and momentum spaces,
\begin{align}
\bm{v} \cdot \frac{\partial f^{m+1} }{\partial \bm{x} } + \bm{F} \cdot  \frac{\partial f^{m+1} }{\partial \bm{p} } =  - \frac{ f^{m+1} -f_{mc}^{eq,m} }{\tau_{mc}^{m} }  - \frac{ f^{m+1} -f_{mr}^{eq,m} }{\tau_{mr}^{m} }, 
\end{align}
where $m$ is the iteration step.
Reformulate above equation into
\begin{align}
\frac{ \Delta f^{m+1} }{\tau_{mc}^{m}} +  \frac{ \Delta f^{m+1} }{ \tau_{mr}^{m} } + \bm{v} \cdot \frac{\partial \Delta f^{m+1} }{\partial \bm{x} } + \bm{F} \cdot  \frac{\partial  \Delta f^{m+1} }{\partial \bm{p} }  =  res^{m},  
\label{eq:residual}
\end{align}
where $\Delta f^{m+1} =f^{m+1}- f^{m}$ is the increment of the distribution function between two adjacent iteration steps and the mesoscopic residual $res$ is defined as
\begin{align}
res^{m} = - \frac{ f^{m} -f_{mc}^{eq,m} }{\tau_{mc}^{m} }  - \frac{ f^{m} -f_{mr}^{eq,m} }{\tau_{mr}^{m} } - \bm{v} \cdot \frac{\partial f^{m} }{\partial \bm{x} } - \bm{F} \cdot  \frac{\partial f^{m} }{\partial \bm{p} }. 
\end{align}
When $res \rightarrow 0$, the iteration converges and $\Delta f \rightarrow 0$.
In other word, the formula of the left hand side of Eq.~\eqref{eq:residual} does not influence the final convergent results.
Therefore, the inexact Newton method is used to solve Eq.~\eqref{eq:residual}~\cite{Chuang17gray}.
First-order upwind scheme is used for the numerical discretization of the left hand side of Eq.~\eqref{eq:residual} and a higher and more accurate numerical discretization method is used for the right.

The whole phase space is discretized into a lot of small pieces, for example, $f_{i,k}$, where $i$ and $k$ the indexes of discretized spatial position and momentum, respectively.
Finite volume method is used to discrete the spatial space
\begin{align}
\bm{v} \cdot \left( \frac{\partial f }{\partial \bm{x} } \right)_i  &=  \frac{1}{V_i} \sum_{j \in N_i } S_{ij}  \bm{v} \cdot \mathbf{n}_{ij} f_{ij},   \\
\bm{v} \cdot \left( \frac{\partial \Delta f }{\partial \bm{x} } \right)_i  &=  \frac{1}{V_i} \sum_{j \in N_i } S_{ij} \bm{v} \cdot \mathbf{n}_{ij}  \Delta f_{ij},  
\end{align}
where $V_i$ is the volume of the cell $i$ in the $\bm{x}$ space, $N(i)$ denotes the sets of face neighbor cell of cell $i$, $ij$ denotes the interface between cell $i$ and cell $j$, $S_{ij}$ is the area of the  interface $ij$, $\mathbf{n}_{ij}$ is the unit normal vector of the interface $ij$ directing from cell $i$ to cell $j$.
For the distribution function at the cell interface,
\begin{align}
\bm{v} \cdot \mathbf{n}_{ij} f_{ij} &= a_1  \left( f_i + \bm{\sigma}_i (\bm{x}_{ij}- \bm{x}_i ) \right)  - a_2 \left( f_j + \bm{\sigma}_j (\bm{x}_{ij}- \bm{x}_j ) \right)  , \\
\bm{v} \cdot \mathbf{n}_{ij}  \Delta f_{ij} &= a_1  \Delta f_i - a_2 \Delta f_j ,   
\end{align}
where $a_1=0.5 \left(  \mathbf{n}_{ij} \cdot \bm{v} + |\mathbf{n}_{ij} \cdot \bm{v} |  \right) $, $a_1=0.5 \left(  \mathbf{n}_{ij} \cdot \bm{v} - |\mathbf{n}_{ij} \cdot \bm{v} |  \right) $, $\bm{\sigma}_i$ is the gradient of distribution function in the cell $i$ calculated by the van Leer limiter, $\bm{x}_{ij}$ is the center of interface $ij$.
Similar strategy is implemented for the numerical discretization of the momentum space,
\begin{align}
\bm{F} \cdot  \left( \frac{\partial f }{\partial \bm{p} } \right)_k  &=  \frac{1}{V_k} \sum_{l \in N_k } S_{kl} \bm{F} \cdot  \mathbf{n}_{kl} f_{kl},   \\
\bm{F} \cdot  \left( \frac{\partial \Delta f }{\partial \bm{p} } \right)_k  &=  \frac{1}{V_k} \sum_{l \in N_k } S_{kl} \bm{F} \cdot  \mathbf{n}_{kl} \Delta f_{kl},
\end{align}
where $V_k$ is the volume of the cell $k$ in the momentum space, $N(k)$ denotes the sets of face neighbor cell of cell $k$, $kl$ denotes the interface between cell $k$ and cell $l$, $S_{kl}$ is the area of the  interface $kl$, $\mathbf{n}_{kl}$ is the unit normal vector of the interface $kl$ directing from cell $k$ to cell $l$.
For the distribution function at the cell interface,
\begin{align}
\bm{F} \cdot  \mathbf{n}_{kl} f_{kl} &= b_1  \left( f_k + \bm{\sigma}_k (\bm{p}_{kl}- \bm{p}_k ) \right)  - b_2 \left( f_l + \bm{\sigma}_l (\bm{p}_{kl}- \bm{p}_l ) \right)  , \\
\bm{F} \cdot  \mathbf{n}_{kl} \Delta f_{kl} &=  b_1 \Delta f_k - b_2  \Delta f_l ,   
\end{align}
where $b_1= 0.5 \left( \mathbf{n}_{kl} \cdot \bm{F} + |\mathbf{n}_{kl} \cdot \bm{F} |  \right)$, $b_2= 0.5 \left( \mathbf{n}_{kl} \cdot \bm{F} - |\mathbf{n}_{kl} \cdot \bm{F} |  \right)$, $\bm{\sigma}_k$ is the gradient of distribution function in the cell $k$ and $\bm{p}_{kl}$ is the center of interface $kl$.
Under the discretized spatial and momentum spaces, Eq.~\eqref{eq:residual} becomes
\begin{align}
&\frac{ \Delta f_{i,k}^{m+1} }{\tau_{i,k,mc}^{m}} +  \frac{ \Delta f_{i,k}^{m+1} }{ \tau_{i,k,mr}^{m} } + \frac{1}{V_i} \sum_{j \in N_i } S_{ij} \bm{v}_k \cdot  \mathbf{n}_{ij}   \Delta f_{ij,k}^{m+1}   + \frac{1}{V_k} \sum_{l \in N_k } S_{kl} \bm{F} \cdot \mathbf{n}_{kl}   \Delta f_{i,kl}^{m+1}   \notag \\
= & - \frac{ f_{i,k}^{m} -f_{i,k,mc}^{eq,m} }{\tau_{i,k,mc}^{m} }  - \frac{ f_{i,k}^{m} -f_{i,k,mr}^{eq,m} }{\tau_{i,k,mr}^{m} }   - \frac{1}{V_i} \sum_{j \in N_i } S_{ij} \bm{v}_k \cdot  \mathbf{n}_{ij} f_{ij,k}^{m} - \frac{1}{V_k} \sum_{l \in N_k } S_{kl} \bm{F} \cdot \mathbf{n}_{kl} f_{i,kl}^{m},  \label{eq:FVMresidual} \\
\Longrightarrow  &  A_{i,k}  \Delta f_{i,k}^{m+1}  = res_{i,k}^{m},
\end{align}
where $A$ is a huge coefficient matrix for the discretized phase space.
The aforementioned large-scale linear equations system is solved by the lower-upper symmetric-Gauss-Seidel method~\cite{YoonS88LUSGS}.

In the hydrodynamic regime, $f \rightarrow f_{mc}^{eq}$, hence such approximation can be used
\begin{align}
\bm{F} \cdot  \frac{\partial f }{\partial \bm{p} }  \approx  \bm{F} \cdot  \frac{\partial f_{mc}^{eq} }{\partial \bm{p} } = f_{mc}^{eq} (f_{mc}^{eq}-1) \bm{F} \cdot   \frac{\bm{v} -\bm{u} }{k_B T}.
\end{align}
Then Eq.~\eqref{eq:FVMresidual} becomes
\begin{align}
&\frac{ \Delta f^{m+1} }{\tau_{mc}^{m}} +  \frac{ \Delta f^{m+1} }{ \tau_{mr}^{m} } + \bm{v} \cdot \frac{\partial \Delta f^{m+1} }{\partial \bm{x} }  =  \frac{ f_{mc}^{eq,m}- f^{m} }{\tau_{mc}^{m} }  - \frac{f_{mr}^{eq,m}-f^{m} }{\tau_{mr}^{m} } - \bm{v} \cdot \frac{\partial f^{m} }{\partial \bm{x} } -  f_{mc}^{eq} (f_{mc}^{eq}-1) \bm{F} \cdot   \frac{\bm{v} -\bm{u} }{k_B T}.  \label{eq:residual_approximation}  \\
&\Longrightarrow  
\frac{ \Delta f_{i,k}^{m+1} }{\tau_{i,k,mc}^{m}} +  \frac{ \Delta f_{i,k}^{m+1} }{ \tau_{i,k,mr}^{m} } + \frac{1}{V_i} \sum_{j \in N_i } S_{ij} \bm{v}_k \cdot  \mathbf{n}_{ij}   \Delta f_{ij,k}^{m+1}   \notag \\
&= - \frac{ f_{i,k}^{m} -f_{i,k,mc}^{eq,m} }{\tau_{i,k,mc}^{m} }  - \frac{ f_{i,k}^{m} -f_{i,k,mr}^{eq,m} }{\tau_{i,k,mr}^{m} }   - \frac{1}{V_i} \sum_{j \in N_i } S_{ij} \bm{v}_k  \cdot  \mathbf{n}_{ij} f_{ij,k}^{m} - f_{i,k,mc}^{eq,m}  (f_{i,k,mc}^{eq,m} -1) \bm{F}  \cdot   \frac{\bm{v}_k -\bm{u}_i }{k_B T_i } .
\end{align}
Above approximations can also be used in the diffusive regime with $f_{mc}^{eq} \rightarrow f_{mr}^{eq}$ and $\bm{u} \rightarrow 0$.
The classical implicit discrete ordinate method is not limited by the electron energy band structure or scattering process.
Electron transport in various materials can be described by this numerical iterative framework as long as the model equation is valid and the input parameters could be obtained regardless of first-principle calculations or empirical formulas.

In this paper, the electron transport and thermal behaviors in single-layer suspended graphene materials are investigated.
The basic physical parameters are as follows.
An isotropic linear dispersion is assumed, i.e., $\bm{k}  = |\bm{k}| \bm{s}$ and $\varepsilon  = \hbar |\bm{k}| v_F$, where $\bm{s}=(\cos \alpha,\sin \alpha)$ is the unit directional vector. 
{\color{blue}{The integral over the momentum space in graphene materials can be reformulated into~\cite{chandra2019a},
\begin{align}
\left< M  \right> = \frac{4}{(2 \pi \hbar)^2 } \int M d\bm{p} = \frac{1}{2 \pi} \int  \int  \frac{2\varepsilon  }{ \pi  (\hbar v_F )^2 }   M d \varepsilon d\Omega.
\end{align}}}
where $d\Omega$ is the integral over the solid angle space.
The doping concentration is $n_D = 10^{12}$ cm$^{-2}$.
Corresponding Fermi velocity and Fermi energy are $v_F=10^6$ m/s~\cite{chandra2019a} and $E_F = \hbar v_F \sqrt{ \pi n_D }$~\cite{PhysRevLett.126.076803}, respectively.
The discrete details of the momentum space are as follows.
The solid angle $\alpha \in [0,2 \pi]$ is discretized equally into $N_{\alpha}=20-40$ pieces in the numerical simulations. 
Electron energy is discretized equally into $N_B=20-40$ pieces in the range $(a,E_F+ N_k k_B T_0)$, where $a$ is the maximum value between $0$ and $E_F - N_k  k_B T_0$, $T_0$ is the background temperature. 
$N_k =10$ in this simulation. 
$M=201$ uniform cells are used for the spatial discretization in each direction, and van Leer limiter~\cite{eDUGKS_2024,vanleer1977} is used for the spatial gradient of distribution function.
{\color{blue}{Numeral validations and grid independence tests have been conducted in Appendix~\ref{sec:validations}.
Results show that the present scheme and discretizations are acceptable to ensure the numerical simulation accuracy.

\subsection{Boundary conditions}

The isothermal boundary condition means that all particles hitting the boundary are absorbed~\cite{eDUGKS_2024}, and particles emitting from the boundary to the computational domain follow the Fermi-Dirac equilibrium distribution with boundary temperature $T_b$ and chemical potential $\mu_b$,
\begin{align}
f({x}_b) = {f}^{eq}_{mr} ( \bm{x}_b, \mu_b, T_b ),\qquad  \bm{s} \cdot \mathbf{n_b} >0,
\end{align}
where $\bm{x}_b$ represents the spatial coordinates of the boundary, $\mathbf{n_b}$ refers to the unit normal vector that pointing from the boundary to the computational domain.
Periodic boundary condition means that when an electron leaves one boundary and meanwhile another electron with the same velocity and energy enters the computational domain from the associated periodic boundary. 
Each electron distribution deviates from its equilibrium state by the same amount at the corresponding periodic boundaries
\begin{align}
 f\left( \bm{x}_{b1},\varepsilon ,\bm{s} \right) -f_{mr}^{eq} \left( T_{b1},\mu _{b1},\varepsilon \right) 
=f\left( \bm{x}_{b2},\varepsilon ,\bm{s} \right) -f_{mr}^{eq} \left( T_{b2},\mu _{b2},\varepsilon \right),
\end{align}
where $b1$ and $b2$ denote the corresponding periodic boundaries, respectively.
Diffusely reflecting boundary condition indicates that the net energy and particle number that penetrate the boundary interface is zero.
Besides electrons reflected from the boundary are isotropic,
\begin{align}
\int_{ \bm{s}\cdot  \mathbf{n_b} < 0 } \bm{v}  f  d \bm{p} &= \int_{ \bm{s}\cdot  \mathbf{n_b} > 0 } \bm{v}  f^{eq}_{mr}(\mu', T')  d \bm{p} ,\\
\int_{ \bm{s}\cdot  \mathbf{n_b} < 0 } \bm{v}  \varepsilon  f  d \bm{p} &= \int_{ \bm{s}\cdot  \mathbf{n_b} > 0 } \bm{v}  \varepsilon  f^{eq}_{mr}(\mu', T')  d \bm{p} ,
\end{align}
The unknown temperature $T'$ and chemical potential $\mu'$ at the boundaries can be obtained by solving above two equations with Newton method.
Specular reflecting boundary condition is
\begin{align}
f\left( \bm{x}_b,\varepsilon,\mu ,\bm{s} \right) =f\left( \bm{x}_b,\varepsilon ,\mu, \bm{s}^{'} \right),  \quad \bm{s}\cdot  \mathbf{n_b} >0
\end{align}
where $\bm{s}^{'}=\bm{s}-2\left( \bm{s}\cdot \mathbf{n_b} \right) \mathbf{n_b}$.

\subsection{Dimensionless Analysis}

Chosen some reference variables including the Fermi velocity $v_F$, characteristic length $L$, background temperature $T_0$ and Fermi energy $E_F$, a dimensionless analysis of eBTE is conducted,
\begin{align}
\bm{s} \cdot \frac{\partial f^* }{\partial \bm{x}^*} +\left( \frac{ -\partial (e \varphi)^* }{\partial \bm{x}^*}  + \bm{s} \times \bm{B}^*  \right)  \cdot \frac{\partial f^* }{\partial \bm{p}^*} = \frac{ f^{eq,*}_{mc} - f^{*} }{  \tau_{mc}^*} +  \frac{ f^{eq,*}_{mr} - f^{*} }{  \tau_{mr}^*},
\label{eq:dimensionlessBTE}
\end{align}
where 
\begin{align}
\tau_{mr}^* &=\frac{\tau_{mr}}{L/v_F}, &\quad  \tau_{mc}^* &=\frac{\tau_{mc}}{L/v_F},  &\quad \bm{x}^* &=\frac{\bm{x}  }{L}   \notag  \\
f^* &=\frac{f}{f_0},  &\quad  (e \varphi )^* &=\frac{ (e \varphi ) }{E_F},  &\quad \bm{B}^* &=\frac{\bm{B}  }{E_F/|e v_F L| }   \notag  \\
\mu^* &=\frac{ \mu }{ E_F},  &\quad  \bm{p}^* &=\frac{ \bm{p} }{ E_F/ v_F }, &\quad T^* &=\frac{ T }{T_0 } \notag  \\
f_{mr}^{eq,*}  &=\frac{f_{mr}^{eq} }{f_0},  &\quad  f_{mc}^{eq,*}  &=\frac{f_{mc}^{eq} }{f_0},  &\quad \bm{u}^* &=\frac{ \bm{u} }{ v_F } .
\end{align}
where $f_0=1/2$.
It can be found that the electron transport is mainly determined by these dimensionless parameters $\tau_{mr}^*$, $\tau_{mc}^*$, $(e \varphi )^*$ and $\bm{B}^*$ for given material properties.
The relationships between the materials size and the length scale of particle transport such as momentum-conserving and momentum-relaxing mean free paths are reflected by dimensionless parameters.}}

\section{Results and discussions}
\label{sec:discussions}

\begin{figure}[htb]
\centering
\includegraphics[width=0.8\textwidth]{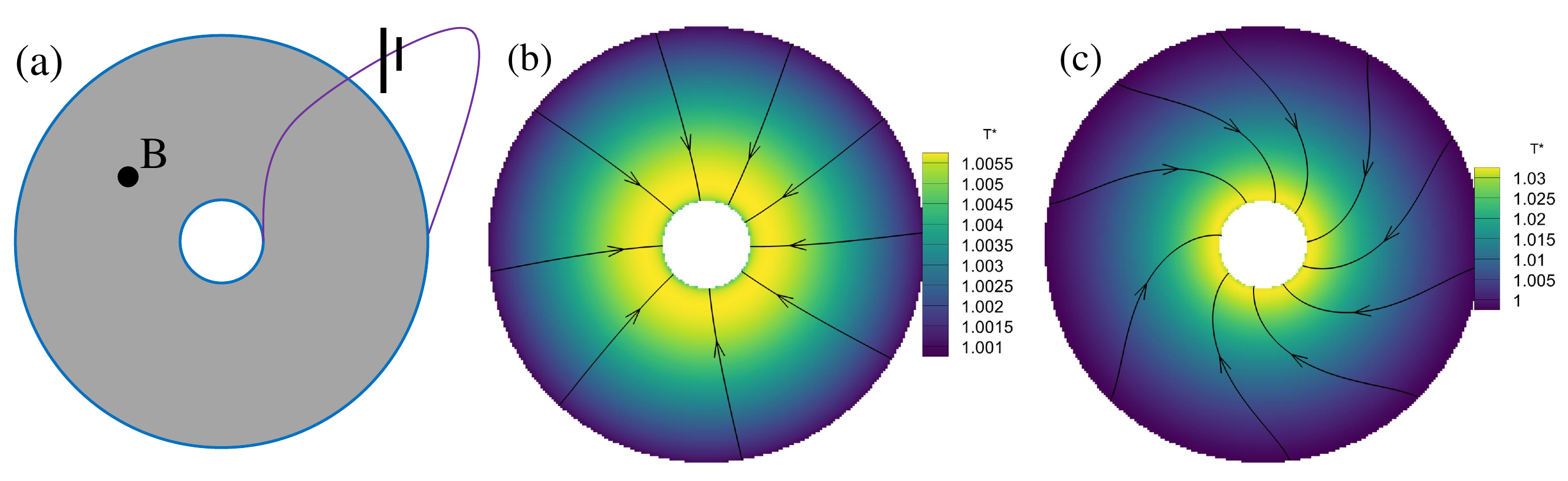}  
\caption{(a) Schematic of a homogeneous Corbino disk geometry driven by electric and magnetic fields. (b,c) Temperature contour and heat flux streamline with a magnetic filed $B^*=0.8560$ perpendicular to the graphene sheet, where (b) $\tau_{mr}^{*}=0.1$, $\tau_{mc}^{*}= \infty$ and (c) $\tau_{mr}^{*}=\infty$, $\tau_{mc}^{*}=0.1$.}
\label{Corbino disk_EM}
\end{figure}

Consider a homogeneous graphene disk~\cite{PhysRevB.105.125302,PhysRevLett.122.137701}, where the ratio of the inner to outer radii of the disk is $1$ to $5$, the thermal effects under the electric and magnetic fields are studied, as shown in~\cref{Corbino disk_EM}(a).
The outer diameter of the disk is regarded as the system characteristic length.
{\color{blue}{Initially the temperature and chemical potential inside the domain are equal to the background temperature and the Fermi energy, respectively.
The initial electron distribution function satisfies the Fermi-Dirac equilibrium distribution.
Temperature and chemical potential in both the inner and outer ends of the disk are fixed at background temperature $T_0=300$ K and Fermi energy, respectively.
Initial particle number density is $N_0$.}}
An electric potential difference $\Delta \varphi$ is applied and a magnetic filed $B$ is applied perpendicular to the graphene disk and pointing out of the paper.

Temperature contour and heat flux streamline in the diffusive and hydrodynamics regimes are plotted in~\cref{Corbino disk_EM}(b,c), respectively, where $(e \Delta \varphi)^*=(e \Delta \varphi)/E_F=0.0856$ and $B^*=0.8560$.
It can be found that the temperature near the inner boundary is higher than that near the outer boundary, and temperature stays a constant tangentially along the radial direction due to symmetry.
Heat flows from outside to inside following the direction of temperature gradient.
Under the same electromagnetic field, the temperature rise in the hydrodynamic regime is larger than that in the diffusive regime.
More importantly, the heat flux flows roughly along the radial direction in the diffusive regime, but there is a certain degree of deflection in the direction of heat flow in the hydrodynamic regime, namely, the heat flux no longer flows only along the radial direction. 
There is heat flux in the tangential direction, corresponding to curved streamlines.

\begin{figure}[htb]
\centering
\includegraphics[width=0.6\textwidth]{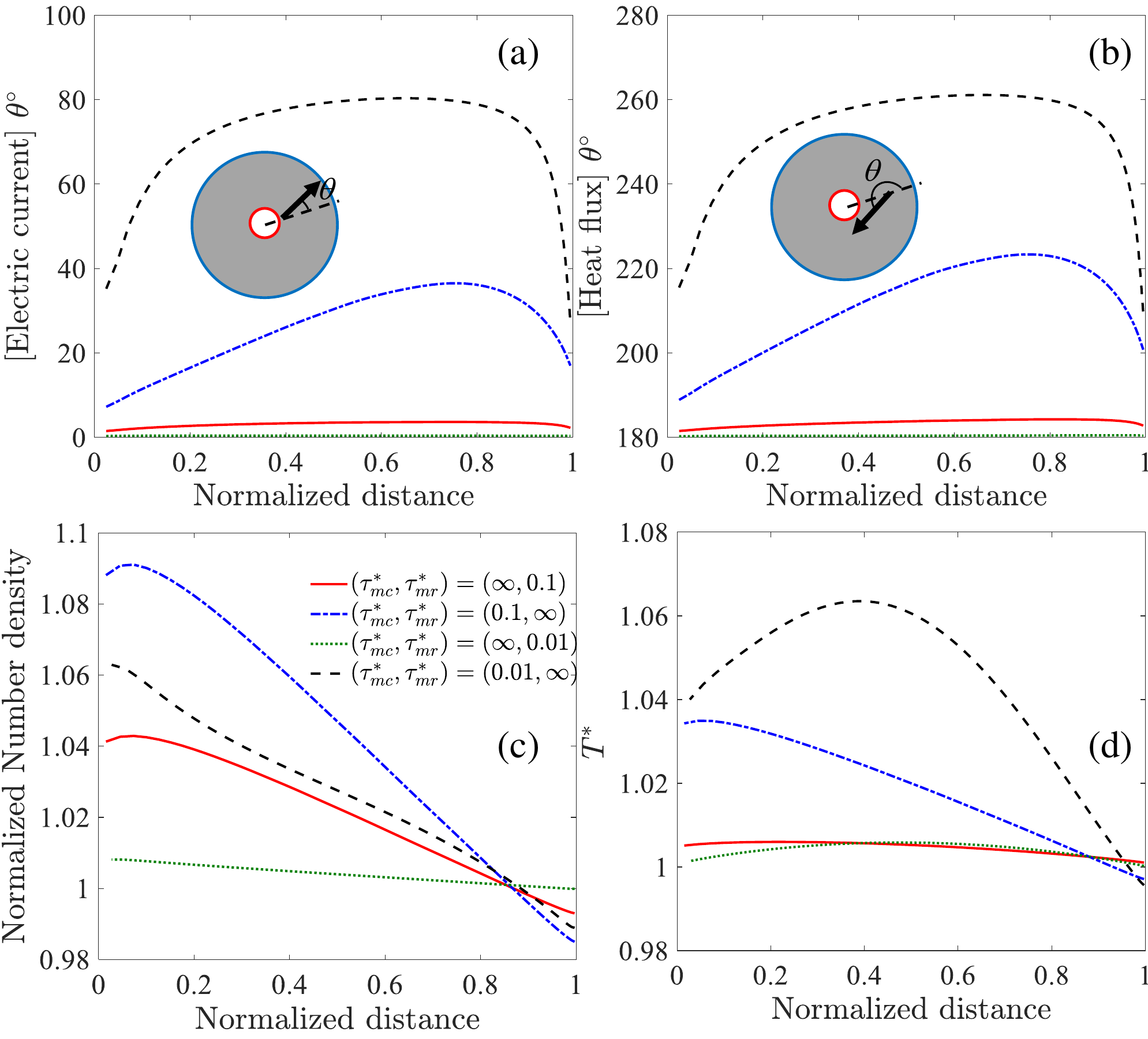 } 
\caption{Macroscopic distributions along the radial direction driven by electric potential gradient and magnetic field $B^*=0.8560$, where normalized distance is $\ln(r/r_{in})/\ln(r_{out}/r_{in})$, $r$ is the distance from the Corbino disk center. Deflection angles of (a) electric current and (b) heat flux. (c) Normalized particle number density $n/N_0$. (d) Normalized temperature.}
\label{electric_driven_GDresults}
\end{figure}

In order to understand the macroscopic distributions in more details, the spatial distributions of particle number density, temperature and the deflection angles $\theta$ of heat flux and electric current along the radial direction from the inside out are plotted in~\cref{electric_driven_GDresults}, where $\tan \theta = q_t/q_r$ or $j_t/j_r$, subscript $t$ and $r$ represents the tangential and radial components of heat flux or electric current, respectively.
{\color{blue}{Please note that if the deflection angle is defined as $\theta= \arctan(q_t/q_r)$ or $\arctan(j_t/j_r)$, it will be difficult to reflect the characteristic that the direction of heat flux is opposite to that of the electric current.}}
Heat flux deflection phenomenon is suppressed by MR scattering process and promoted by MC scattering process, as shown in~\cref{electric_driven_GDresults}(a)(b). 
Electrons are constantly moving from the outside to the inside under the electric field forces, which causes electrons to accumulate near the inner circle.
As the electrons move inward, they feel the Lorentz forces from the magnetic field and interact with other electrons.
{\color{blue}{The radial temperature gradient is established by the competition between Joule heating (which is stronger in the inner region where the current density is higher) and thermoelectric cooling (or Peltier effects). 
In the hydrodynamic regime, reduced momentum relaxation leads to lower dissipation and hence a larger temperature rise compared to the diffusive case. 
The heat flux, contains both a dissipative Fourier-like component (opposing the temperature gradient) and a thermoelectric component (driven by the electric current). 
In the present simulations, the latter dominates in the hydrodynamic regime, resulting in heat flow from the outer (cooler) boundary toward the inner (hotter) boundary. 
This is a classic thermoelectric response and does not violate Fourier's law.}}
\begin{figure}[htb]
\centering
\includegraphics[width=0.80\textwidth]{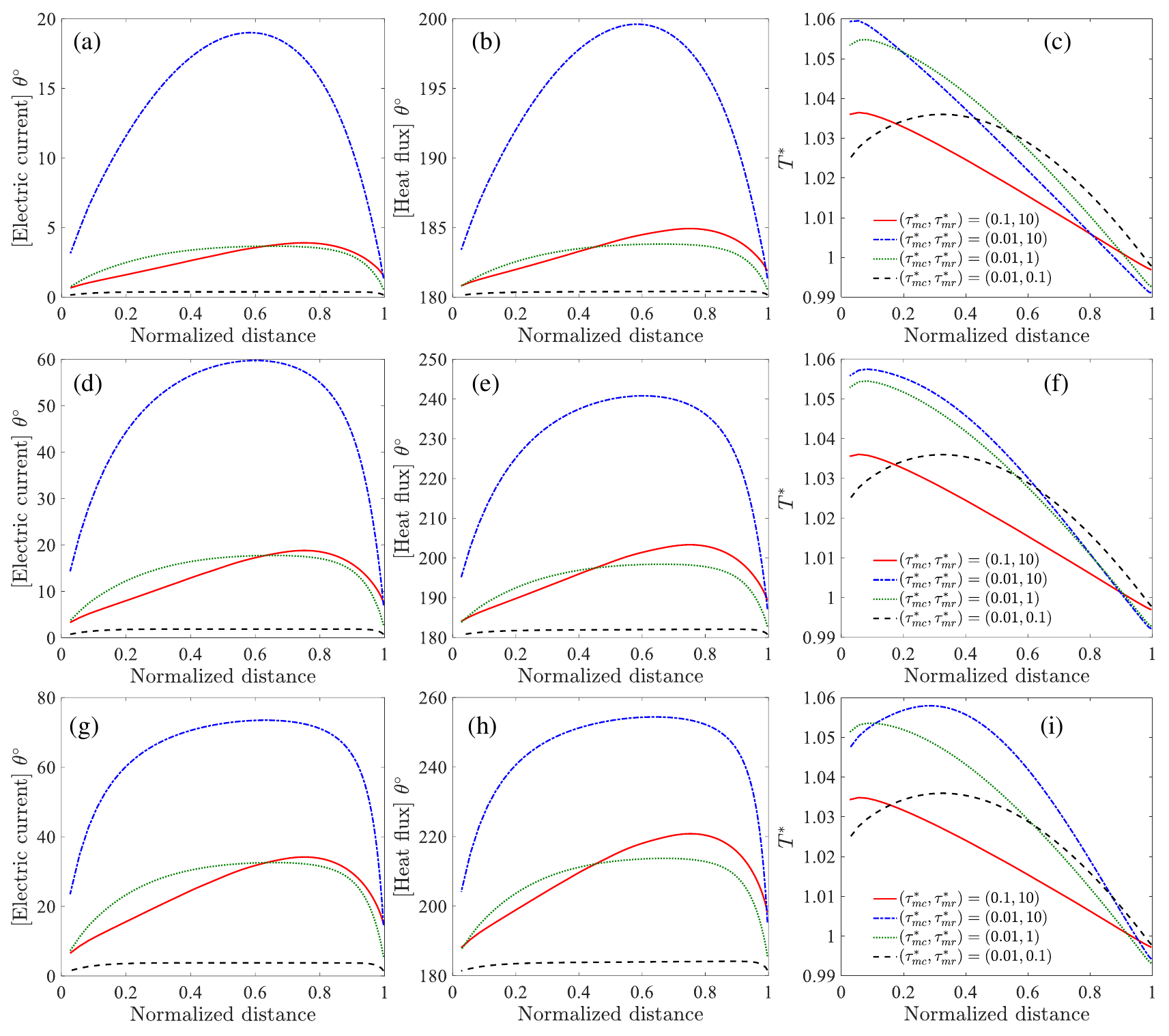}
\caption{Macroscopic distributions along the radial direction driven by electric potential gradient and various magnetic field, where normalized distance is $\ln(r/r_{in})/\ln(r_{out}/r_{in})$, $r$ is the distance from the Corbino disk center. The magnetic field in the first, second, third row is $B^* = 0.0856$, $B^* = 0.428$, $B^* = 0.856$, respectively. Deflection angles of (a,d,g) electric current and (b,e,h) heat flux. (c,f,i) Normalized temperature.}
\label{electric_force_results}
\end{figure}

In the diffusive regime, sufficient MR scattering process happens and the frequent energy and momentum exchange among electrons drives the distribution function into a local Fermi-Dirac distribution.
On one hand, frequent MR scattering process prevents electrons from constantly moving from the outside in and gathering around the inner circle, so that the difference in particle number density between the inside and the outside becomes smaller.
On the other hand, the acceleration tendency of electrons under the action of electric field forces is also destroyed by frequent MR scattering process. 
Specifically, although the electron is accelerated by the electric field force to obtain the corresponding momentum increment in a mean free path range, this momentum increment is immediately consumed by the MR scattering, which results in a small momentum or energy increment of the electron so that the temperature rise in the whole region is small.
{\color{blue}{When $\tau_{mr}^* \rightarrow 0$ and $\tau_{mc}^* \rightarrow \infty$, a first-order Chapman-Enskog expansion of distribution function leads to 
\begin{align}
f^*  &= f^{eq,*}_{mr} -\tau_{mr}^*  \bm{s} \cdot \frac{\partial f^{eq,*}_{mr} }{\partial \bm{x}^* } + \tau_{mr}^*  \frac{\partial (e  \varphi^*)}{\partial \bm{x}^*} \cdot  \frac{\partial f^{eq,*}_{mr} }{\partial \bm{p}^* }  
-  \tau_{mr}^*   \left( \bm{s} \times \bm{B}^*  \right)  \cdot  \frac{\partial f^{eq,*}_{mr} }{\partial \bm{p}^* }  + o \left( \tau_{mr}^{*,2} \right)   \notag \\
\Longrightarrow  \left< \bm{s} f^* \right>&\approx -\tau_{mr}^* \left<  \bm{s} \bm{s} \cdot \left( \frac{\partial f^{eq,*}_{mr} }{\partial T^*}  \frac{\partial T^*}{\partial \bm{x}^* }  + \frac{\partial f^{eq,*}_{mr} }{\partial \mu^*}  \frac{\partial \mu^*}{\partial \bm{x}^* } \right)  \right> +   \tau_{mr}^*  \left< \bm{s}  \frac{\partial (e  \varphi^*)}{\partial \bm{x}^*} \cdot  \frac{\partial f^{eq,*}_{mr} }{\partial  \varepsilon^* } \frac{\partial  \varepsilon^*}{\partial \bm{p}^* }     \right>  \notag \\
 &\approx  -\tau_{mr}^* \left<  \bm{s} \bm{s} \cdot \left( \frac{\partial f^{eq,*}_{mr} }{\partial T^*}  \frac{\partial T^*}{\partial \bm{x}^* }  + \frac{\partial f^{eq,*}_{mr} }{\partial \mu^*}  \frac{\partial \mu^*}{\partial \bm{x}^* } \right)  \right>  +  \tau_{mr}^*  \left< \bm{s}  \frac{\partial (e  \varphi^*)}{\partial \bm{x}^*} \cdot  \left(  -\frac{\partial f^{eq,*}_{mr} }{\partial  \mu^* } \right)  \bm{s}     \right>   \notag \\
 &\approx  -\tau_{mr}^* \left<  \bm{s} \bm{s} \cdot   \frac{\partial f^{eq,*}_{mr} }{\partial T^*}  \right>    \frac{\partial T^*}{\partial \bm{x}^* }  +  \tau_{mr}^* \left<  - \bm{s} \bm{s}   \frac{\partial f^{eq,*}_{mr} }{\partial \mu^*}    \right> \cdot   \left( \frac{\partial \mu^*}{\partial \bm{x}^* } +  \frac{\partial (e  \varphi^*)}{\partial \bm{x}^*}  \right),  \label{eq:CEMR}
\end{align}
The first-order magnetic term disappears due to $ \left( \bm{s} \times \bm{B^*}  \right)   \cdot  \frac{\partial f^{eq,*}_{mr} }{\partial \bm{p}^* }  =0 $.}}
Note that the gradients of temperature, chemical potential and electric potential are along the radial direction due to symmetry.
Hence the deflection of macroscopic flux~\eqref{eq:CEMR} tangentially along the radial direction in the diffusive regime is much small and on the order of $o \left( \tau_{mr}^{*,2} \right)  $ regardless of heat flux or electric current.

In the hydrodynamic regime, sufficient MC scattering process results in a local equilibrium state with a nonzero drift velocity $\bm{u}^*$.
Frequent MC scattering process results in a much smaller resistance, which makes it easier for electrons to flow from the outside in so that the particle number density inside increases significantly, as shown in~\cref{electric_driven_GDresults}(c). 
On the other hand, small resistance indicates small dissipation and larger temperature rise under the same electric potential difference, as shown in~\cref{electric_driven_GDresults}(d).
{\color{blue}{When $\tau_{mr}^* \rightarrow \infty$ and $\tau_{mc}^* \rightarrow 0$, the first-order Chapman-Enskog expansion of distribution function gives  
\begin{align}
f^* &= f^{eq,*}_{mc} -\tau_{mc}^*  \bm{s} \cdot \frac{\partial f^{eq,*}_{mc} }{\partial \bm{x}^* } - \tau_{mc}^*   \left(-\frac{\partial (e  \varphi^*)}{\partial \bm{x}^*}  + \bm{s} \times \bm{B}^*  \right)  \cdot  \frac{\partial f^{eq,*}_{mc} }{\partial \bm{p}^* } + o \left( \tau_{mc}^{*,2} \right)  \notag \\
\Longrightarrow  \left<  \bm{s}f^* \right> &\approx \left<  \bm{s} f^{eq,*}_{mc} \right>  -\tau_{mc}^*  \left<  \bm{s} \bm{s} \cdot \left( \frac{\partial f^{eq,*}_{mc} }{\partial T^*}  \frac{\partial T^*}{\partial \bm{x}^* } +\frac{\partial f^{eq,*}_{mc} }{\partial \mu^*}  \frac{\partial \mu^*}{\partial \bm{x}^* } +\frac{\partial f^{eq,*}_{mc} }{\partial \bm{u}^*}  \cdot \frac{\partial \bm{u}^*}{\partial \bm{x}^* }  \right)  \right>   \notag \\
& - \left< \tau_{mc}^* \bm{s}   \left(-\frac{\partial (e  \varphi^*)}{\partial \bm{x}^*} +   \bm{s} \times \bm{B}^*  \right)  \cdot  \frac{\partial f^{eq,*}_{mc} }{\partial \varepsilon^* } \left( \frac{ \partial \varepsilon^* }{\partial \bm{p}^* } - \bm{u}^*   \right)   \right>  \notag \\
&\approx \left<  \bm{s} f^{eq,*}_{mc} \right>  + \tau_{mc}^*  \left<  -\bm{s} \bm{s} \cdot  \frac{\partial f^{eq,*}_{mc} }{\partial T^*}    \right>  \frac{\partial T^*}{\partial \bm{x}^* }+  \tau_{mc}^* \left<  - \bm{s} \bm{s}   \frac{\partial f^{eq,*}_{mc} }{\partial \mu^*}    \right> \cdot   \left( \frac{\partial \mu^*}{\partial \bm{x}^* } +  \frac{\partial (e  \varphi^*)}{\partial \bm{x}^*}  \right) \notag \\
&+  \tau_{mc}^* \left< - \frac{\partial f^{eq,*}_{mc} }{\partial \mu^* } \bm{s}   \left(-\frac{\partial (e  \varphi^*)}{\partial \bm{x}^*} +   \bm{s} \times \bm{B}^*  \right)   \right>\cdot  \bm{u} +  \tau_{mc}^*  \left< - \frac{\partial f^{eq,*}_{mc} }{\partial \mu^* } \bm{s} \bm{s} \cdot \bm{p}^* \right> \cdot \frac{\partial \bm{u}^*}{\partial \bm{x}^* } 
\end{align}
It can be found that nonzero drift velocity is significantly affected by the electromagnetic field.
The Chapmen-Enskog expansion analysis reveals that the flux deflection originates from the emergence of a collective drift velocity in the electron distribution function in the hydrodynamic limit. 
This electron drift, subject to the Lorentz force, acquires a tangential component in disk system that cannot be compensated by electric forces in the bulk region. 
Consequently, the flux deflection is a genuine bulk phenomenon and is largely insensitive to the detailed choice of boundary conditions.}}

{\color{blue}{Numerical results within various magnetic filed force, MR/MC scattering process are shown in~\cref{electric_force_results}.
It can be found that the deflection angle decreases with the magnetic field force under the same scattering rates.
The deflection angle increases when the MC scattering process increases but decreases when the MR scattering process increases.}}
Based on dimensionless analysis of eBTE~\eqref{eq:dimensionlessBTE}, it can be found that the local flux in the tangential direction of radius depends on the temperature gradient, electric and magnetic field force and the local electron scattering strength.
When MR scattering dominates, there are large momentum relaxation so that the acceleration of electrons by the electric field is less effective.
The electron mobility is reduced, and consequently the effect of magnetic field force is naturally reduced, making it difficult to change the direction of electron motion.
Hence, electron motion in the tangential direction tends to zero, and the deflection phenomenon is weak.
When MC scattering process dominates, there are few momentum relaxation and the electric field force drives electron to move from outside in.
During this motion, it feels a tangential Lorentz force.
Different from the Hall effect in cuboid geometry, the Lorentz force in disk geometry is usually not in the radial direction.
Therefore, it is not easy to be compensated by the electrostatic field force which is always in the radial direction due to the symmetry of particle number density.
Consequently, the macroscopic flux deflection phenomenon appears.

\begin{figure}[htb]
\centering
\includegraphics[width=0.8\textwidth]{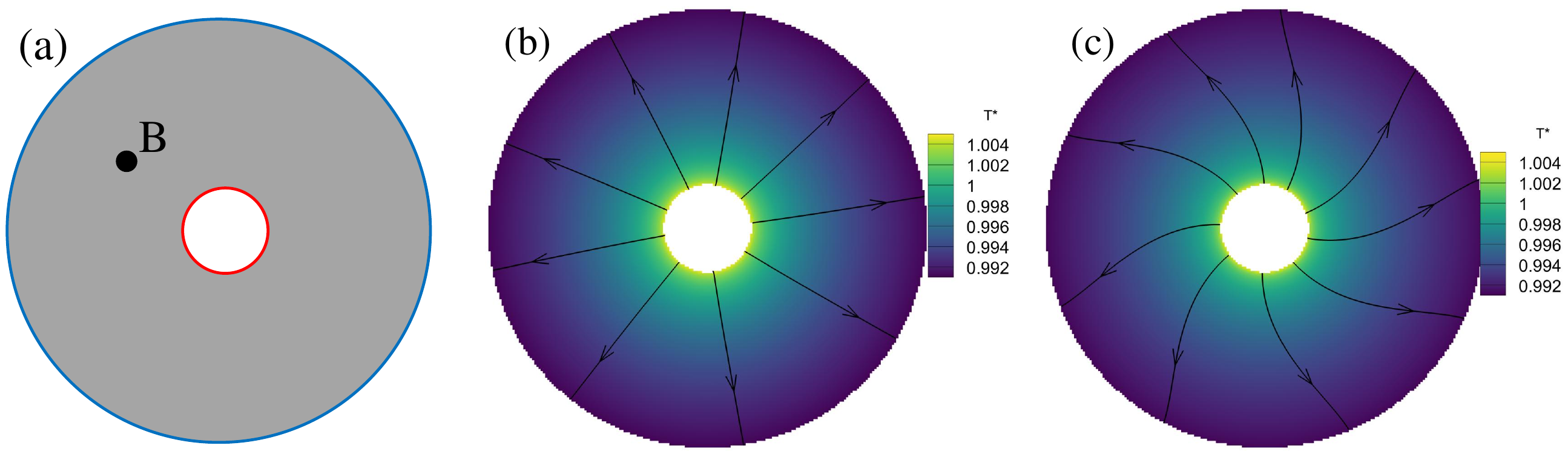}  
\caption{(a) Schematic of a homogeneous Corbino disk geometry driven by temperature gradients and magnetic fields. (b,c) Temperature contour and heat flux streamline with a magnetic filed $B^*=0.8560$ perpendicular to the graphene sheet, where (b) $\tau_{mr}^{*}=0.1$, $\tau_{mc}^{*}= \infty$ and (c) $\tau_{mr}^{*}=\infty$, $\tau_{mc}^{*}=0.1$.}
\label{Corbino disk_TM}
\end{figure}

Secondly, the thermal effects in graphene Corbino disk under the temperature gradient and magnetic field are studied, as shown in~\cref{Corbino disk_TM}(a). 
In this case, the electric and chemical potentials in the inner and outer ends of the disk are the same.
The dimensionless temperature $T^*$ in the inner and outer boundaries is $1.01$ and $0.99$, respectively.
\begin{figure}[htb]
\centering
\includegraphics[width=0.6\textwidth]{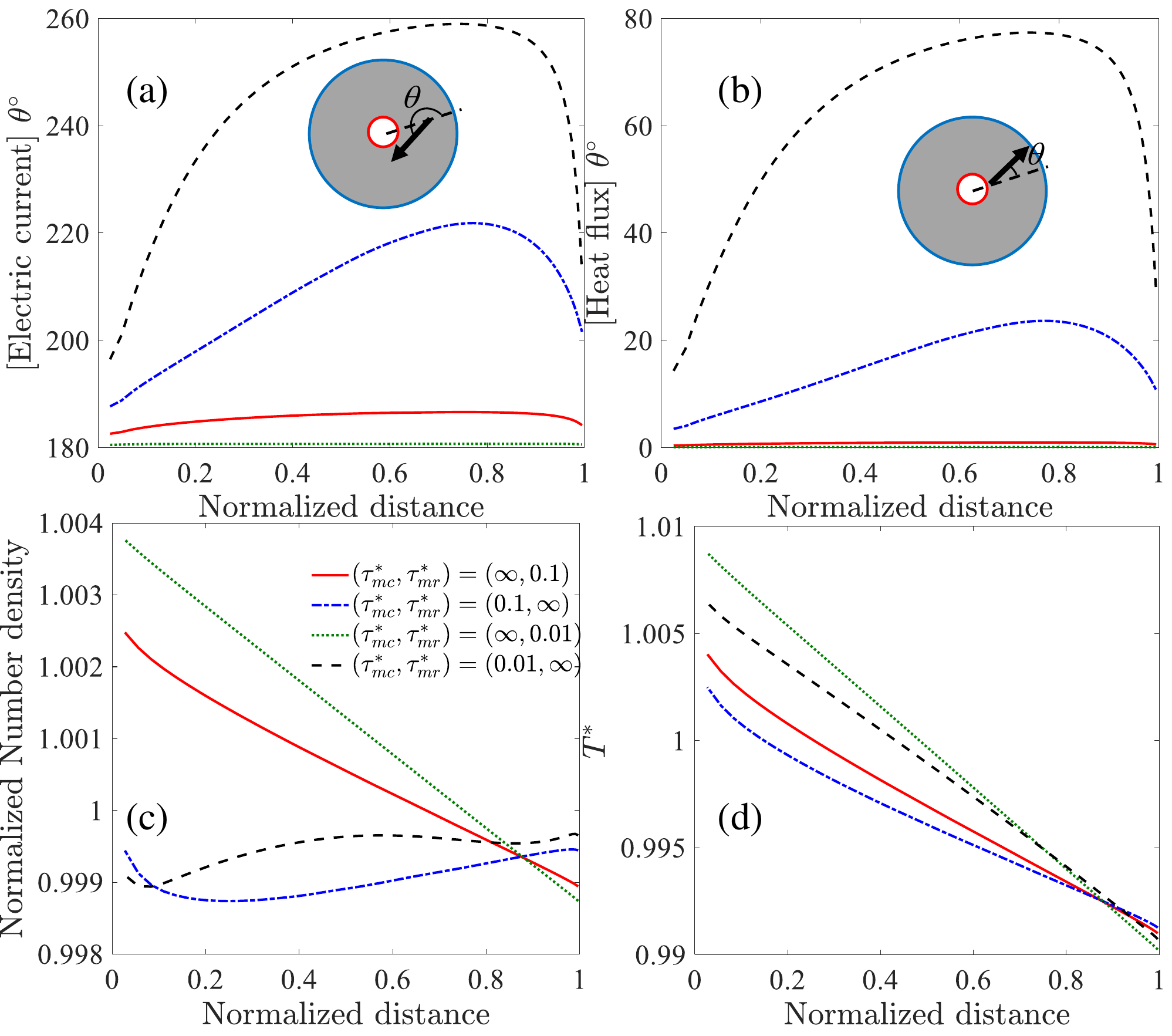 } 
\caption{Macroscopic distributions along the radial direction driven by temperature gradient and magnetic field $B^*=0.8560$, where normalized distance is $\ln(r/r_{in})/\ln(r_{out}/r_{in})$, $r$ is the distance from the disk center. Deflection angles of (a) electric current and (b) heat flux. (c) Normalized particle number density $n/N_0$. (d) Normalized temperature.}
\label{temperature_driven_GDresults}
\end{figure}
\begin{figure}[htb]
\centering
\includegraphics[width=0.8\textwidth]{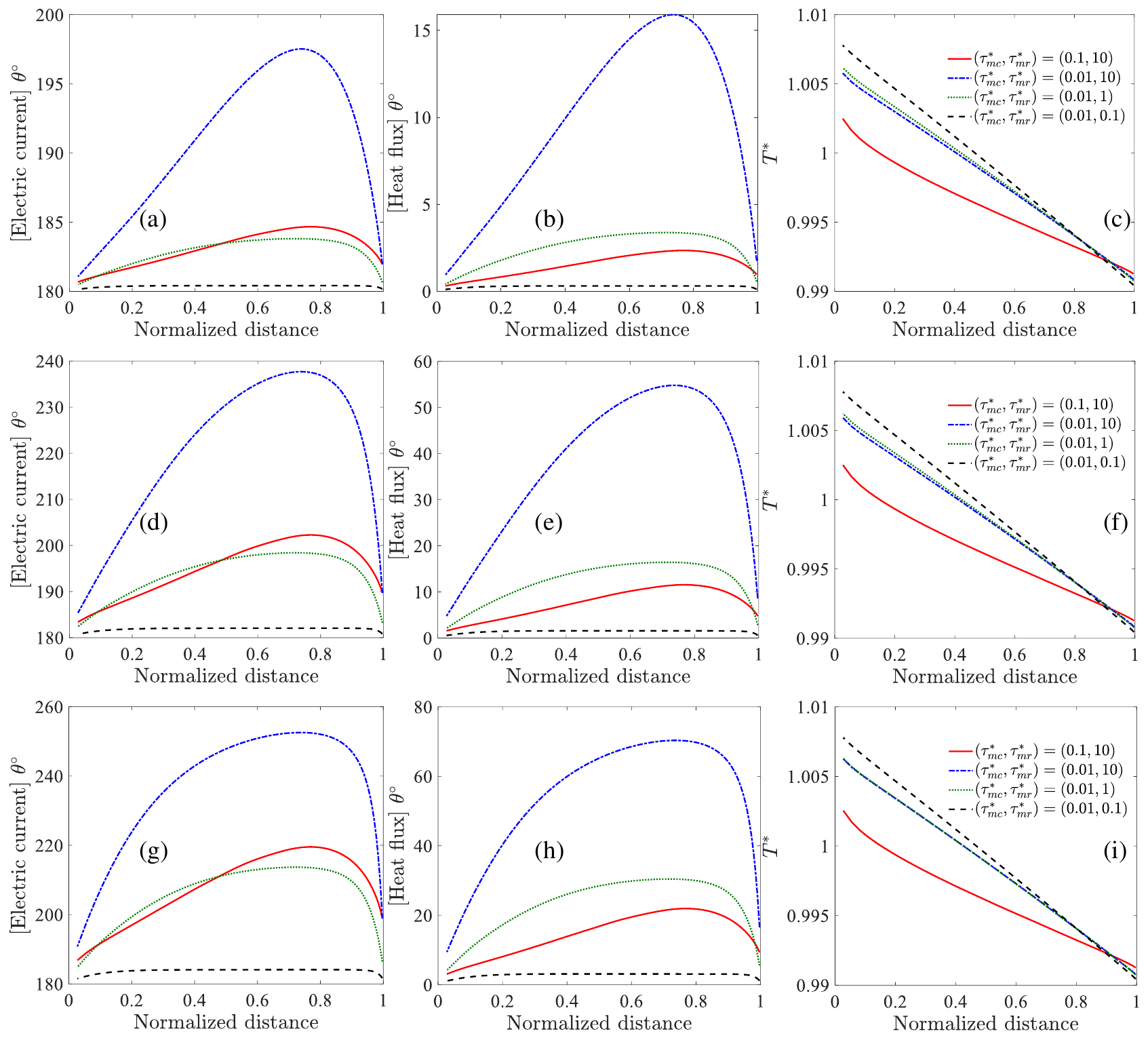}
\caption{Macroscopic distributions along the radial direction driven by temperature gradient and various magnetic field, where normalized distance is $\ln(r/r_{in})/\ln(r_{out}/r_{in})$, $r$ is the distance from the Corbino disk center. The magnetic field in the first, second, third row is $B^* = 0.0856$, $B^* = 0.428$, $B^* = 0.856$, respectively. Deflection angles of (a,d,g) electric current and (b,e,h) heat flux. (c,f,i) Normalized temperature.}
\label{temperature_force_results}
\end{figure}

Numerical results are shown in~\cref{Corbino disk_TM,temperature_driven_GDresults,temperature_force_results}.
The particle number density changes slightly inside the disk due to small temperature variance, which indicates that the electric field force is much small.
It can be found that there are larger temperature slip near the inner boundary than that outside, because the inner radius is smaller so that the ballistic effects inside is more serious.
Similarly there is little heat flux deflection in the diffusive regime due to the sufficient MR scattering process and the heat flux defection is obvious in the hydrodynamic regime.
Opposite to the above results shown in~\cref{Corbino disk_EM}(b,c) driven by electromagnetic fields, the heat flux flows from inside out and electric current flows from outside in under the temperature gradient.

We elucidate the underlying physical mechanisms in term of the evolution of electron distribution function in the momentum space.
In the hydrodynamic regime, electron transport presents collective behavior, namely, electron distribution function is generally deviated in a certain direction with drift velocity $\bm{u}^*$ in the momentum space. 
When electric field is applied, electron suffers from a radial outside-in electric field force, which causes its distribution function to deviate from equilibrium state towards the center of the disk. Therefore, the electron will also be driven by a clockwise magnetic field force to move clockwise under the action of a magnetic field, which leads to that the electric current flows anticlockwise and heat flux flows clockwise.
Different from electric field force, the temperature gradient drives the electron distribution function to deviate from equilibrium in a direction away from the center of the disk. 
Therefore, the electron will be driven by a counterclockwise magnetic field force to move counterclockwise under the action of a magnetic field, which leads to that the electric current flows clockwise and heat flux flows anticlockwise.

{\color{blue}{In a word,~\cref{electric_driven_GDresults,electric_force_results}, ~\cref{temperature_driven_GDresults,temperature_force_results} systematically examine the magnetic-field-tuned electron transport in a Corbino disk under electric-field and temperature-gradient drives, revealing the essential distinction between hydrodynamic and diffusive regimes in their magnetic response. The deflection angle $\theta$ increases monotonically with $B^*$, yet its magnitude is governed by the dominant scattering mechanism: enhanced MC scattering (reduced $\tau_{mc}^*$) sustains a collective drift that enables the Lorentz force to generate a persistent transverse momentum, yielding a pronounced deflection as a hallmark of hydrodynamics; enhanced MR scattering (reduced $\tau_{mr}^*$) dissipates momentum within each mean free path, quenching the Lorentz accumulation and restoring diffusive transport with $\theta \rightarrow 0$. The deflection chirality reverses upon switching the drive — inward drift under electric field produces clockwise deflection, while outward diffusion under temperature gradient flips the Lorentz force direction. 
Thermally, the hydrodynamic regime under electric drive exhibits a substantial temperature rise due to suppressed dissipation, whereas the diffusive regime remains nearly isothermal.
Under temperature-gradient drive, the temperature rise is weak and scattering-insensitive, as the heat flux is boundary-controlled rather than dissipation-dominated. 
Collectively, the deflection amplitude and temperature rise serve as independent diagnostics for hydrodynamic versus diffusive transport, displaying clear scaling separation under the competition between scattering rates and magnetic field.}}

\section{Conclusion}
\label{sec:conclusions}

{\color{blue}{A kinetic simulation of electron thermal and charge transport in a graphene Corbino disk has been performed, with the magnetic field strength and MC/MR scattering rates being systematically scanned under the framework of BTE. 
Three physical distinctions between the hydrodynamic and diffusive regimes are identified. 
Firstly, flux deflection by the Lorentz force is found to be a hydrodynamic signature, pronounced when electron–electron scattering dominates and quenched when momentum-relaxing scattering prevails. 
In the diffusive limit, the first-order magnetic correction to the flux vanishes by symmetry, while in the hydrodynamic limit, the finite drift velocity allows the Lorentz force to generate a sustained tangential component.
Secondly, a significant temperature rise is produced in the hydrodynamic regime under electric-field driving due to reduced energy dissipation, whereas a weaker thermal response is induced under temperature-gradient driving, which is less sensitive to the electron MC/MR scattering rates. 
Thirdly, the deflection chirality under temperature-gradient driving is observed to be opposite to that under electric-field driving. 
The current results indicate that thermodynamic physical quantities, especially the heat flow deflection caused by magnetic fields, can also be used as an approach to distinguish hydrodynamic electron flow and classical diffusive transport. 
However, the spatially resolved, high-resolution measurement~\cite{zhangx_2026_thermoX,yao2026imaging,beardo2026phonon} of heat flux or temperature is much more difficult than that of electric current or potential. 
Perhaps with the development of future experimental techniques, it will be possible to detect the spatial distribution of heat flux.}}

\section*{Author Statements}

\textbf{Chuang Zhang}: Supervision, Conceptualization, Investigation, Methodology, Numerical analysis, Funding acquisition, Writing - original draft.
\textbf{Meng Lian}: Investigation, Methodology, Numerical analysis, Writing-review \& editing.
\textbf{Hong Liang}: Numerical analysis, Writing-review \& editing.
\textbf{Xiaokang Li}: Numerical analysis, Writing-review \& editing.
\textbf{Zhaoli Guo}: Conceptualization, Methodology, Numerical analysis, Writing-review \& editing.
\textbf{Jing-Tao L\"u}: Supervision, Conceptualization, Investigation, Numerical analysis, Writing-review \& editing.

\section*{Conflict of interest}

Jing-Tao L\"u is a member of the editorial board of Thermo-X.

\section*{Acknowledgments}

C. Z. acknowledges the support of the National Natural Science Foundation of China (Grant No. 52506078) and the Zhejiang Provincial Natural Science Foundation of China under Grant No. LMS26E060012. 
The authors acknowledge Beijing PARATERA Tech CO., Ltd. for the HPC resources.

\section*{APPENDIX: Numerical Validations and Grid Independence}
\label{sec:validations}

{\color{blue}{Modular program verification is a commonly used method in large-scale computing. 
This appendix will gradually verify the numerical performance of the present algorithm.

\begin{figure}[htb]
\centering
\includegraphics[width=0.6\textwidth]{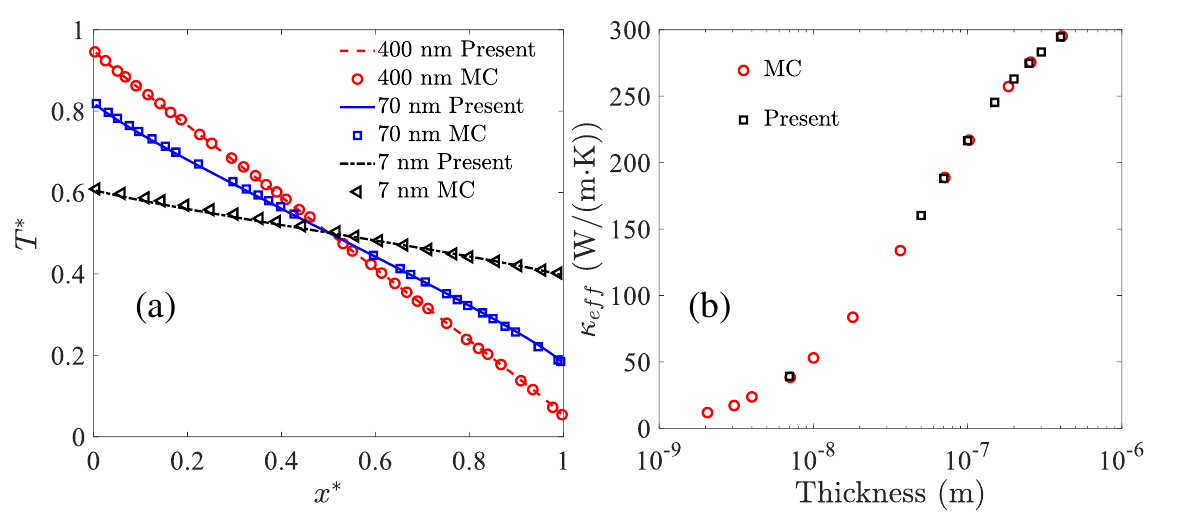}
\caption{(a) Steady-state temperature spatial distributions, where $x^*=x/L$ and $T^*=(T-T_R)/(T_L-T_R)$ are the normalized coordinate and temperature, respectively. (b) Size effects of cross-plane electron thermal transport in Au films, where $\kappa_{eff}= qL/(T_L-T_R)$. `MC' is the data obtained from reference~\cite{PhysRevB.99.205433} and `Present' is the present results. }
\label{1Dtest}
\end{figure}
\begin{figure}[htb]
\centering
\includegraphics[width=0.6\textwidth]{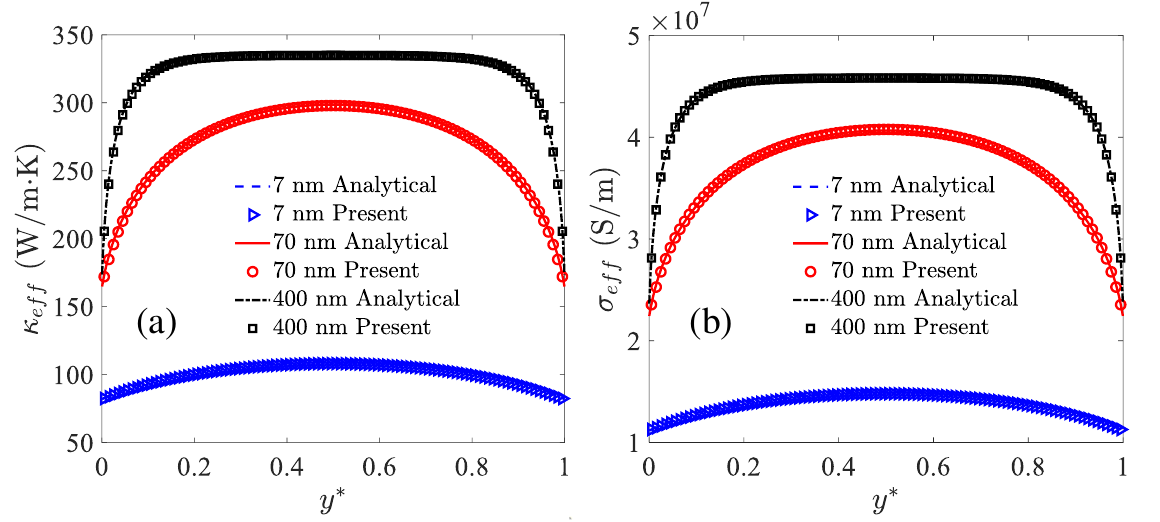}
\caption{(a) Spatial distributions of the effective thermal conductivity under the temperature gradient. (b) Spatial distributions of the effective electric conductivity under the electric filed force. `Analytical' is the analytical solutions obtained from previous papers~\cite{Fuchs_1938_analytical,Sondheimer_1952_analytical,eDUGKS_2024} and `Present' is the present results. $\kappa_{eff}= q/(dT/dx)$, $\sigma_{eff}=j/(E)$, $y^*=y/H$ is the normalized coordinate.}
\label{2Dtest}
\end{figure}
Firstly, the cross-plane electronic heat conduction in a thin Au metal film is simulated with thickness $L$.
The temperatures at both ends of the film are fixed with $T_L=T_0+\Delta T$, $T_R=T_0-\Delta T$, respectively, where $\Delta T \ll T_0$.
Isothermal boundary condition is applied for both ends of the film.
The electronic band structure of the Au material is treated using the free electron approximation model.
Specific electronic property parameters and energy band structures of the Au materials can be found in our previous paper~\cite{eDUGKS_2024} as well as the numerical discretizations.
The initial chemical potential of the entire computational domain is all at the Fermi level $E_F=5.51$ eV.
Numerical results predicted by the present scheme are compared with the Monte Carlo solutions~\cite{PhysRevB.99.205433} at different film thickness, as shown in~\cref{1Dtest}.
The profiles showed that the present scheme can accurately capture different-sized electronic heat conduction phenomena under the temperature gradients.

Secondly, the in-plane electron transport in Au materials is simulated under temperature gradient or electric field force.
Diffusely reflecting adiabatic or insulating boundary conditions are adopted for both the upper and lower boundaries, and periodic boundary conditions are used for left and right boundaries.
The distance between the upper and lower boundaries is $H$.
When a constant and very small temperature gradient is applied in the left-right direction, the system temperature increase is not significant, and the chemical potential inside is approximately the Fermi energy level. 
A parabolic-like heat flux distribution will be generated in the vertical direction of the temperature gradient and an analytical solution can be obtained for this electronic thermal transport process.
Similarly, when a constant and very small electric filed force is applied in the left-right direction, the system temperature increase is not significant.  
A parabolic-like electronic current distribution will be generated in the vertical direction of the electric field force and an analytical solution can be obtained for this electron transport process~\cite{Fuchs_1938_analytical,Sondheimer_1952_analytical,eDUGKS_2024}.
It can be found that the numerical results predicted by the present scheme are in excellent agreement with the analytical solutions at different $H$ in~\cref{2Dtest}.

Thirdly, grid independence validation is conducted for graphene Corbino disk under the electric field, as shown in~\cref{electric_driven_GDresults}(a), where $N_{\alpha}$, $N_B$ and $M$ are the number of discrete solid angle, energy bands and spatial cells in each direction.
When $\tau_{mr}^* =0.01$ or $\tau_{mc}^* =0.01$, 
`Less mesh': $N_{\alpha}=20$, $N_B=20$, $M=200$, $N_k =10$, 
`More mesh': $N_{\alpha}=40$, $N_B=40$, $M=300$, $N_k =15$.
When $\tau_{mr}^* =0.1$ or $\tau_{mc}^* =0.1$, 
`Less mesh': $N_{\alpha}=40$, $N_B=40$, $M=200$, $N_k =10$, 
`More mesh': $N_{\alpha}=80$, $N_B=96$, $M=300$, $N_k =15$.
It can be found that the present numerical discretizations of the phase space are acceptable to ensure the numerical simulation accuracy.
\begin{figure}[htb]
\centering
\includegraphics[width=0.8\textwidth]{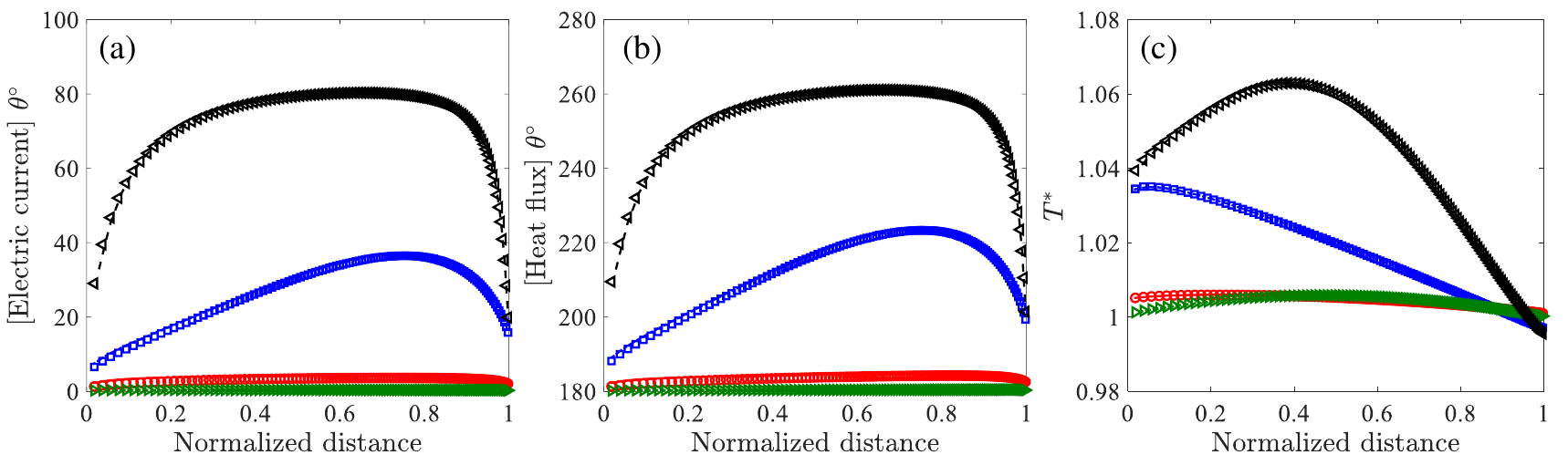}
\caption{All introduction are the same as~\cref{electric_driven_GDresults}, where lines ares the `Less mesh', and symbols are `More mesh'.}
\label{gridvalidation}
\end{figure}
}}

\bibliographystyle{elsarticle-num-names_clear}

\bibliography{phonon}

\begin{thebibliography}{71}
\expandafter\ifx\csname natexlab\endcsname\relax\def\natexlab#1{#1}\fi
\providecommand{\url}[1]{\texttt{#1}}
\providecommand{\href}[2]{#2}
\providecommand{\path}[1]{#1}
\providecommand{\DOIprefix}{doi:}
\providecommand{\ArXivprefix}{arXiv:}
\providecommand{\URLprefix}{URL: }
\providecommand{\Pubmedprefix}{pmid:}
\providecommand{\doi}[1]{\href{http://dx.doi.org/#1}{\path{#1}}}
\providecommand{\Pubmed}[1]{\href{pmid:#1}{\path{#1}}}
\providecommand{\bibinfo}[2]{#2}
\ifx\xfnm\relax \def\xfnm[#1]{\unskip,\space#1}\fi
%Type = Article
\bibitem[{Varnavides et~al.(2023)Varnavides, Yacoby, Felser, and
  Narang}]{hydrodynamic_2023charge_review}
\bibinfo{author}{G.~Varnavides}, \bibinfo{author}{A.~Yacoby},
  \bibinfo{author}{C.~Felser}, \bibinfo{author}{P.~Narang},
\newblock \bibinfo{title}{Charge transport and hydrodynamics in materials},
\newblock \bibinfo{journal}{Nature Reviews Materials} \bibinfo{volume}{8}
  (\bibinfo{year}{2023}) \bibinfo{pages}{726--741}. \URLprefix
  \url{https://doi.org/10.1038/s41578-023-00597-3}.
  \DOIprefix\doi{10.1038/s41578-023-00597-3}.
%Type = Article
\bibitem[{Fritz and Scaffidi(2024)}]{hydrodynamic_electron_review}
\bibinfo{author}{L.~Fritz}, \bibinfo{author}{T.~Scaffidi},
\newblock \bibinfo{title}{Hydrodynamic electronic transport},
\newblock \bibinfo{journal}{Annual Review of Condensed Matter Physics}
  \bibinfo{volume}{15} (\bibinfo{year}{2024}) \bibinfo{pages}{17--44}.
  \URLprefix \url{https://doi.org/10.1146/annurev-conmatphys-040521-042014}.
  \DOIprefix\doi{10.1146/annurev-conmatphys-040521-042014}.
%Type = Article
\bibitem[{Wolf et~al.(2023)Wolf, Aharon-Steinberg, Yan, and
  Holder}]{para_hydrodynamics_flakes_2023}
\bibinfo{author}{Y.~Wolf}, \bibinfo{author}{A.~Aharon-Steinberg},
  \bibinfo{author}{B.~Yan}, \bibinfo{author}{T.~Holder},
\newblock \bibinfo{title}{Para-hydrodynamics from weak surface scattering in
  ultraclean thin flakes},
\newblock \bibinfo{journal}{Nat. Commun.} \bibinfo{volume}{14}
  (\bibinfo{year}{2023}) \bibinfo{pages}{2334}. \URLprefix
  \url{https://www.nature.com/articles/s41467-023-37966-z}.
  \DOIprefix\doi{10.1038/s41467-023-37966-z}.
%Type = Article
\bibitem[{Beardo et~al.(2026)Beardo, Tur-Prats, Camacho, and
  Alvarez}]{beardo2026phonon}
\bibinfo{author}{A.~Beardo}, \bibinfo{author}{J.~Tur-Prats},
  \bibinfo{author}{J.~Camacho}, \bibinfo{author}{F.~X. Alvarez},
\newblock \bibinfo{title}{Phonon hydrodynamics: Theory and experiments},
\newblock \bibinfo{journal}{Thermo-X} \bibinfo{volume}{2}
  (\bibinfo{year}{2026}) \bibinfo{pages}{202513}. \URLprefix
  \url{https://doi.org/10.70401/tx.2026.0008}.
%Type = Article
\bibitem[{Lian et~al.(2024)Lian, Zhang, Guo, and L\"u}]{eDUGKS_2024}
\bibinfo{author}{M.~Lian}, \bibinfo{author}{C.~Zhang},
  \bibinfo{author}{Z.~Guo}, \bibinfo{author}{J.-T. L\"u},
\newblock \bibinfo{title}{Discrete unified gas kinetic scheme for the solution
  of electron boltzmann transport equation with callaway approximation},
\newblock \bibinfo{journal}{Phys. Rev. E} \bibinfo{volume}{109}
  (\bibinfo{year}{2024}) \bibinfo{pages}{065310}. \URLprefix
  \url{https://link.aps.org/doi/10.1103/PhysRevE.109.065310}.
  \DOIprefix\doi{10.1103/PhysRevE.109.065310}.
%Type = Article
\bibitem[{Zhu and L{\"u}(2025)}]{zhu2025out}
\bibinfo{author}{G.-L. Zhu}, \bibinfo{author}{J.-T. L{\"u}},
\newblock \bibinfo{title}{Out-of-equilibrium ultrafast electron and phonon
  energy transfer dynamics in metals: The role of non-thermal effect},
\newblock \bibinfo{journal}{Thermo-X} \bibinfo{volume}{2}
  (\bibinfo{year}{2025}) \bibinfo{pages}{202512}. \URLprefix
  \url{https://doi.org/10.70401/tx.2025.0007}.
%Type = Article
\bibitem[{Chandra et~al.(2019)Chandra, Kataria, Sahdev, and
  Sundararaman}]{chandra2019a}
\bibinfo{author}{M.~Chandra}, \bibinfo{author}{G.~Kataria},
  \bibinfo{author}{D.~Sahdev}, \bibinfo{author}{R.~Sundararaman},
\newblock \bibinfo{title}{Hydrodynamic and ballistic {{AC}} transport in
  two-dimensional {{Fermi}} liquids},
\newblock \bibinfo{journal}{Phys. Rev. B} \bibinfo{volume}{99}
  (\bibinfo{year}{2019}) \bibinfo{pages}{165409}. \URLprefix
  \url{https://link.aps.org/doi/10.1103/PhysRevB.99.165409}.
  \DOIprefix\doi{10.1103/PhysRevB.99.165409}.
%Type = Article
\bibitem[{Guo et~al.(2017)Guo, Ilseven, Falkovich, and
  Levitov}]{pnas_superballistic2017}
\bibinfo{author}{H.~Guo}, \bibinfo{author}{E.~Ilseven},
  \bibinfo{author}{G.~Falkovich}, \bibinfo{author}{L.~S. Levitov},
\newblock \bibinfo{title}{Higher-than-ballistic conduction of viscous electron
  flows},
\newblock \bibinfo{journal}{Proceedings of the National Academy of Sciences}
  \bibinfo{volume}{114} (\bibinfo{year}{2017}) \bibinfo{pages}{3068--3073}.
  \URLprefix \url{https://www.pnas.org/doi/abs/10.1073/pnas.1612181114}.
  \DOIprefix\doi{10.1073/pnas.1612181114}.
%Type = Article
\bibitem[{Sulpizio et~al.(2019)Sulpizio, Ella, Rozen, Birkbeck, Perello, Dutta,
  Ben-Shalom, Taniguchi, Watanabe, Holder et~al.}]{sulpizio2019visualizing}
\bibinfo{author}{J.~A. Sulpizio}, \bibinfo{author}{L.~Ella},
  \bibinfo{author}{A.~Rozen}, \bibinfo{author}{J.~Birkbeck},
  \bibinfo{author}{D.~J. Perello}, \bibinfo{author}{D.~Dutta},
  \bibinfo{author}{M.~Ben-Shalom}, \bibinfo{author}{T.~Taniguchi},
  \bibinfo{author}{K.~Watanabe}, \bibinfo{author}{T.~Holder}, et~al.,
\newblock \bibinfo{title}{Visualizing poiseuille flow of hydrodynamic
  electrons},
\newblock \bibinfo{journal}{Nature} \bibinfo{volume}{576}
  (\bibinfo{year}{2019}) \bibinfo{pages}{75--79}. \URLprefix
  \url{https://www.nature.com/articles/s41586-019-1788-9}.
  \DOIprefix\doi{https://doi.org/10.1038/s41586-019-1788-9}.
%Type = Article
\bibitem[{Sano and Matsuo(2023)}]{PhysRevLett.130.166201}
\bibinfo{author}{R.~Sano}, \bibinfo{author}{M.~Matsuo},
\newblock \bibinfo{title}{Breaking down the magnonic wiedemann-franz law in the
  hydrodynamic regime},
\newblock \bibinfo{journal}{Phys. Rev. Lett.} \bibinfo{volume}{130}
  (\bibinfo{year}{2023}) \bibinfo{pages}{166201}. \URLprefix
  \url{https://link.aps.org/doi/10.1103/PhysRevLett.130.166201}.
  \DOIprefix\doi{10.1103/PhysRevLett.130.166201}.
%Type = Article
\bibitem[{Li et~al.(2022)Li, Andreev, and Levchenko}]{PhysRevB.105.155307}
\bibinfo{author}{S.~Li}, \bibinfo{author}{A.~V. Andreev},
  \bibinfo{author}{A.~Levchenko},
\newblock \bibinfo{title}{Hydrodynamic electron transport in graphene hall-bar
  devices},
\newblock \bibinfo{journal}{Phys. Rev. B} \bibinfo{volume}{105}
  (\bibinfo{year}{2022}) \bibinfo{pages}{155307}. \URLprefix
  \url{https://link.aps.org/doi/10.1103/PhysRevB.105.155307}.
  \DOIprefix\doi{10.1103/PhysRevB.105.155307}.
%Type = Article
\bibitem[{Shavit et~al.(2019)Shavit, Shytov, and
  Falkovich}]{PhysRevLett.123.026801}
\bibinfo{author}{M.~Shavit}, \bibinfo{author}{A.~Shytov},
  \bibinfo{author}{G.~Falkovich},
\newblock \bibinfo{title}{Freely flowing currents and electric field expulsion
  in viscous electronics},
\newblock \bibinfo{journal}{Phys. Rev. Lett.} \bibinfo{volume}{123}
  (\bibinfo{year}{2019}) \bibinfo{pages}{026801}. \URLprefix
  \url{https://link.aps.org/doi/10.1103/PhysRevLett.123.026801}.
  \DOIprefix\doi{10.1103/PhysRevLett.123.026801}.
%Type = Article
\bibitem[{Gurzhi(1963)}]{Gurzhi_1963}
\bibinfo{author}{R.~N. Gurzhi},
\newblock \bibinfo{title}{Minimum of resistance in impurity-free conductors},
\newblock \bibinfo{journal}{Sov. Phys. JETP} \bibinfo{volume}{17}
  (\bibinfo{year}{1963}) \bibinfo{pages}{521}. \URLprefix
  \url{http://jetp.ras.ru/cgi-bin/e/index/e/17/2/p521?a=list}.
%Type = Article
\bibitem[{Gurzhi(1968)}]{Gurzhi_1968}
\bibinfo{author}{R.~N. Gurzhi},
\newblock \bibinfo{title}{Hydrodynamic effects in solids and at low
  temperature},
\newblock \bibinfo{journal}{Sov. Phys.-Usp.} \bibinfo{volume}{11}
  (\bibinfo{year}{1968}) \bibinfo{pages}{255--270}. \URLprefix
  \url{https://doi.org/10.1070%2Fpu1968v011n02abeh003815}.
  \DOIprefix\doi{10.1070/pu1968v011n02abeh003815}.
%Type = Article
\bibitem[{Black(1980)}]{PhysRevB.21.3279}
\bibinfo{author}{J.~E. Black},
\newblock \bibinfo{title}{Contribution of electron-electron normal scattering
  processes to the electrical resistivity of thin wires},
\newblock \bibinfo{journal}{Phys. Rev. B} \bibinfo{volume}{21}
  (\bibinfo{year}{1980}) \bibinfo{pages}{3279--3286}. \URLprefix
  \url{https://link.aps.org/doi/10.1103/PhysRevB.21.3279}.
  \DOIprefix\doi{10.1103/PhysRevB.21.3279}.
%Type = Article
\bibitem[{de~Jong and Molenkamp(1995)}]{PhysRevB.51.13389}
\bibinfo{author}{M.~J.~M. de~Jong}, \bibinfo{author}{L.~W. Molenkamp},
\newblock \bibinfo{title}{Hydrodynamic electron flow in high-mobility wires},
\newblock \bibinfo{journal}{Phys. Rev. B} \bibinfo{volume}{51}
  (\bibinfo{year}{1995}) \bibinfo{pages}{13389--13402}. \URLprefix
  \url{https://link.aps.org/doi/10.1103/PhysRevB.51.13389}.
  \DOIprefix\doi{10.1103/PhysRevB.51.13389}.
%Type = Article
\bibitem[{Gennaro and Rettori(1984)}]{Gennaro_1984}
\bibinfo{author}{S.~D. Gennaro}, \bibinfo{author}{A.~Rettori},
\newblock \bibinfo{title}{The low-temperature electrical resistivity of
  potassium: size effects and the role of normal electron-electron scattering},
\newblock \bibinfo{journal}{Journal of Physics F: Metal Physics}
  \bibinfo{volume}{14} (\bibinfo{year}{1984}) \bibinfo{pages}{L237--L242}.
  \URLprefix \url{https://doi.org/10.1088/0305-4608/14/12/001}.
  \DOIprefix\doi{10.1088/0305-4608/14/12/001}.
%Type = Article
\bibitem[{Gurzhi et~al.(1995)Gurzhi, Kalinenko, and
  Kopeliovich}]{PhysRevLett.74.3872}
\bibinfo{author}{R.~N. Gurzhi}, \bibinfo{author}{A.~N. Kalinenko},
  \bibinfo{author}{A.~I. Kopeliovich},
\newblock \bibinfo{title}{Electron-electron collisions and a new hydrodynamic
  effect in two-dimensional electron gas},
\newblock \bibinfo{journal}{Phys. Rev. Lett.} \bibinfo{volume}{74}
  (\bibinfo{year}{1995}) \bibinfo{pages}{3872--3875}. \URLprefix
  \url{https://link.aps.org/doi/10.1103/PhysRevLett.74.3872}.
  \DOIprefix\doi{10.1103/PhysRevLett.74.3872}.
%Type = Article
\bibitem[{Yu et~al.(1984)Yu, Haerle, Zwart, Bass, Pratt, and
  Schroeder}]{PhysRevLett.52.368}
\bibinfo{author}{Z.~Z. Yu}, \bibinfo{author}{M.~Haerle}, \bibinfo{author}{J.~W.
  Zwart}, \bibinfo{author}{J.~Bass}, \bibinfo{author}{W.~P. Pratt},
  \bibinfo{author}{P.~A. Schroeder},
\newblock \bibinfo{title}{Negative temperature derivative of resistivity in
  thin potassium samples: The gurzhi effect?},
\newblock \bibinfo{journal}{Phys. Rev. Lett.} \bibinfo{volume}{52}
  (\bibinfo{year}{1984}) \bibinfo{pages}{368--371}. \URLprefix
  \url{https://link.aps.org/doi/10.1103/PhysRevLett.52.368}.
  \DOIprefix\doi{10.1103/PhysRevLett.52.368}.
%Type = Article
\bibitem[{Ella et~al.(2019)Ella, Rozen, Birkbeck, {Ben-Shalom}, Perello,
  Zultak, Taniguchi, Watanabe, Geim, Ilani, and Sulpizio}]{ella2019}
\bibinfo{author}{L.~Ella}, \bibinfo{author}{A.~Rozen},
  \bibinfo{author}{J.~Birkbeck}, \bibinfo{author}{M.~{Ben-Shalom}},
  \bibinfo{author}{D.~Perello}, \bibinfo{author}{J.~Zultak},
  \bibinfo{author}{T.~Taniguchi}, \bibinfo{author}{K.~Watanabe},
  \bibinfo{author}{A.~K. Geim}, \bibinfo{author}{S.~Ilani},
  \bibinfo{author}{J.~A. Sulpizio},
\newblock \bibinfo{title}{Simultaneous voltage and current density imaging of
  flowing electrons in two dimensions},
\newblock \bibinfo{journal}{Nat. Nanotechnol.} \bibinfo{volume}{14}
  (\bibinfo{year}{2019}) \bibinfo{pages}{480--487}. \URLprefix
  \url{https://www.nature.com/articles/s41565-019-0398-x}.
  \DOIprefix\doi{10.1038/s41565-019-0398-x}.
%Type = Article
\bibitem[{Tomadin et~al.(2014)Tomadin, Vignale, and
  Polini}]{PhysRevLett.113.235901}
\bibinfo{author}{A.~Tomadin}, \bibinfo{author}{G.~Vignale},
  \bibinfo{author}{M.~Polini},
\newblock \bibinfo{title}{Corbino disk viscometer for 2d quantum electron
  liquids},
\newblock \bibinfo{journal}{Phys. Rev. Lett.} \bibinfo{volume}{113}
  (\bibinfo{year}{2014}) \bibinfo{pages}{235901}. \URLprefix
  \url{https://link.aps.org/doi/10.1103/PhysRevLett.113.235901}.
  \DOIprefix\doi{10.1103/PhysRevLett.113.235901}.
%Type = Article
\bibitem[{Briskot et~al.(2015)Briskot, Sch\"utt, Gornyi, Titov, Narozhny, and
  Mirlin}]{PhysRevB.92.115426}
\bibinfo{author}{U.~Briskot}, \bibinfo{author}{M.~Sch\"utt},
  \bibinfo{author}{I.~V. Gornyi}, \bibinfo{author}{M.~Titov},
  \bibinfo{author}{B.~N. Narozhny}, \bibinfo{author}{A.~D. Mirlin},
\newblock \bibinfo{title}{Collision-dominated nonlinear hydrodynamics in
  graphene},
\newblock \bibinfo{journal}{Phys. Rev. B} \bibinfo{volume}{92}
  (\bibinfo{year}{2015}) \bibinfo{pages}{115426}. \URLprefix
  \url{https://link.aps.org/doi/10.1103/PhysRevB.92.115426}.
  \DOIprefix\doi{10.1103/PhysRevB.92.115426}.
%Type = Article
\bibitem[{Zeng et~al.(2019)Zeng, Li, Dietrich, Ghosh, Watanabe, Taniguchi,
  Hone, and Dean}]{PhysRevLett.122.137701}
\bibinfo{author}{Y.~Zeng}, \bibinfo{author}{J.~I.~A. Li},
  \bibinfo{author}{S.~A. Dietrich}, \bibinfo{author}{O.~M. Ghosh},
  \bibinfo{author}{K.~Watanabe}, \bibinfo{author}{T.~Taniguchi},
  \bibinfo{author}{J.~Hone}, \bibinfo{author}{C.~R. Dean},
\newblock \bibinfo{title}{High-quality magnetotransport in graphene using the
  edge-free corbino geometry},
\newblock \bibinfo{journal}{Phys. Rev. Lett.} \bibinfo{volume}{122}
  (\bibinfo{year}{2019}) \bibinfo{pages}{137701}. \URLprefix
  \url{https://link.aps.org/doi/10.1103/PhysRevLett.122.137701}.
  \DOIprefix\doi{10.1103/PhysRevLett.122.137701}.
%Type = Article
\bibitem[{Gabbana et~al.(2018)Gabbana, Polini, Succi, Tripiccione, and
  Pellegrino}]{PhysRevLett.121.236602}
\bibinfo{author}{A.~Gabbana}, \bibinfo{author}{M.~Polini},
  \bibinfo{author}{S.~Succi}, \bibinfo{author}{R.~Tripiccione},
  \bibinfo{author}{F.~M.~D. Pellegrino},
\newblock \bibinfo{title}{Prospects for the detection of electronic
  preturbulence in graphene},
\newblock \bibinfo{journal}{Phys. Rev. Lett.} \bibinfo{volume}{121}
  (\bibinfo{year}{2018}) \bibinfo{pages}{236602}. \URLprefix
  \url{https://link.aps.org/doi/10.1103/PhysRevLett.121.236602}.
  \DOIprefix\doi{10.1103/PhysRevLett.121.236602}.
%Type = Article
\bibitem[{Kumar et~al.(2022)Kumar, Birkbeck, Sulpizio, Perello, Taniguchi,
  Watanabe, Reuven, Scaffidi, Stern, Geim, and
  Ilani}]{nature_2022_hydrodynamics}
\bibinfo{author}{C.~Kumar}, \bibinfo{author}{J.~Birkbeck},
  \bibinfo{author}{J.~A. Sulpizio}, \bibinfo{author}{D.~Perello},
  \bibinfo{author}{T.~Taniguchi}, \bibinfo{author}{K.~Watanabe},
  \bibinfo{author}{O.~Reuven}, \bibinfo{author}{T.~Scaffidi},
  \bibinfo{author}{A.~Stern}, \bibinfo{author}{A.~K. Geim},
  \bibinfo{author}{S.~Ilani},
\newblock \bibinfo{title}{Imaging hydrodynamic electrons flowing without
  landauer-sharvin resistance},
\newblock \bibinfo{journal}{Nature} \bibinfo{volume}{609}
  (\bibinfo{year}{2022}) \bibinfo{pages}{276–281}.
  \DOIprefix\doi{https://doi.org/10.1038/s41586-022-05002-7}.
%Type = Article
\bibitem[{Vool et~al.(2021)Vool, Hamo, Varnavides, Wang, Zhou, Kumar,
  Dovzhenko, Qiu, Garcia, Pierce et~al.}]{vool2021imaging}
\bibinfo{author}{U.~Vool}, \bibinfo{author}{A.~Hamo},
  \bibinfo{author}{G.~Varnavides}, \bibinfo{author}{Y.~Wang},
  \bibinfo{author}{T.~X. Zhou}, \bibinfo{author}{N.~Kumar},
  \bibinfo{author}{Y.~Dovzhenko}, \bibinfo{author}{Z.~Qiu},
  \bibinfo{author}{C.~A. Garcia}, \bibinfo{author}{A.~T. Pierce}, et~al.,
\newblock \bibinfo{title}{Imaging phonon-mediated hydrodynamic flow in wte2},
\newblock \bibinfo{journal}{Nat. Phys.} \bibinfo{volume}{17}
  (\bibinfo{year}{2021}) \bibinfo{pages}{1216--1220}.
  \DOIprefix\doi{https://doi.org/10.1038/s41567-021-01341-w}.
%Type = Article
\bibitem[{Jaoui et~al.(2021)Jaoui, Fauqu{\'e}, and Behnia}]{jaoui2021thermal}
\bibinfo{author}{A.~Jaoui}, \bibinfo{author}{B.~Fauqu{\'e}},
  \bibinfo{author}{K.~Behnia},
\newblock \bibinfo{title}{Thermal resistivity and hydrodynamics of the
  degenerate electron fluid in antimony},
\newblock \bibinfo{journal}{Nat. Commun.} \bibinfo{volume}{12}
  (\bibinfo{year}{2021}) \bibinfo{pages}{195}.
  \DOIprefix\doi{https://doi.org/10.1038/s41467-020-20420-9}.
%Type = Article
\bibitem[{Krebs et~al.(2023)Krebs, Behn, Li, Smith, Watanabe, Taniguchi,
  Levchenko, and Brar}]{viscous_2023_hydrodynamics}
\bibinfo{author}{Z.~J. Krebs}, \bibinfo{author}{W.~A. Behn},
  \bibinfo{author}{S.~Li}, \bibinfo{author}{K.~J. Smith},
  \bibinfo{author}{K.~Watanabe}, \bibinfo{author}{T.~Taniguchi},
  \bibinfo{author}{A.~Levchenko}, \bibinfo{author}{V.~W. Brar},
\newblock \bibinfo{title}{Imaging the breaking of electrostatic dams in
  graphene for ballistic and viscous fluids},
\newblock \bibinfo{journal}{Science} \bibinfo{volume}{379}
  (\bibinfo{year}{2023}) \bibinfo{pages}{671--676}. \URLprefix
  \url{https://www.science.org/doi/abs/10.1126/science.abm6073}.
  \DOIprefix\doi{10.1126/science.abm6073}.
%Type = Article
\bibitem[{Ku et~al.(2020)Ku, Zhou, Li, Shin, Shi, Burch, Anderson, Pierce, Xie,
  Hamo et~al.}]{ku2020imaging}
\bibinfo{author}{M.~J. Ku}, \bibinfo{author}{T.~X. Zhou},
  \bibinfo{author}{Q.~Li}, \bibinfo{author}{Y.~J. Shin}, \bibinfo{author}{J.~K.
  Shi}, \bibinfo{author}{C.~Burch}, \bibinfo{author}{L.~E. Anderson},
  \bibinfo{author}{A.~T. Pierce}, \bibinfo{author}{Y.~Xie},
  \bibinfo{author}{A.~Hamo}, et~al.,
\newblock \bibinfo{title}{Imaging viscous flow of the dirac fluid in graphene},
\newblock \bibinfo{journal}{Nature} \bibinfo{volume}{583}
  (\bibinfo{year}{2020}) \bibinfo{pages}{537--541}. \URLprefix
  \url{https://www.nature.com/articles/s41586-020-2507-2}.
  \DOIprefix\doi{https://doi.org/10.1038/s41586-020-2507-2}.
%Type = Article
\bibitem[{Stern et~al.(2022)Stern, Scaffidi, Reuven, Kumar, Birkbeck, and
  Ilani}]{PhysRevLett.129.157701}
\bibinfo{author}{A.~Stern}, \bibinfo{author}{T.~Scaffidi},
  \bibinfo{author}{O.~Reuven}, \bibinfo{author}{C.~Kumar},
  \bibinfo{author}{J.~Birkbeck}, \bibinfo{author}{S.~Ilani},
\newblock \bibinfo{title}{How electron hydrodynamics can eliminate the
  landauer-sharvin resistance},
\newblock \bibinfo{journal}{Phys. Rev. Lett.} \bibinfo{volume}{129}
  (\bibinfo{year}{2022}) \bibinfo{pages}{157701}. \URLprefix
  \url{https://link.aps.org/doi/10.1103/PhysRevLett.129.157701}.
  \DOIprefix\doi{10.1103/PhysRevLett.129.157701}.
%Type = Article
\bibitem[{Bandurin et~al.(2018)Bandurin, Shytov, Levitov, Kumar, Berdyugin,
  Shalom, Grigorieva, Geim, and Falkovich}]{bandurin2018}
\bibinfo{author}{D.~A. Bandurin}, \bibinfo{author}{A.~V. Shytov},
  \bibinfo{author}{L.~S. Levitov}, \bibinfo{author}{R.~K. Kumar},
  \bibinfo{author}{A.~I. Berdyugin}, \bibinfo{author}{M.~B. Shalom},
  \bibinfo{author}{I.~V. Grigorieva}, \bibinfo{author}{A.~K. Geim},
  \bibinfo{author}{G.~Falkovich},
\newblock \bibinfo{title}{Fluidity onset in graphene},
\newblock \bibinfo{journal}{Nat. Commun.} \bibinfo{volume}{9}
  (\bibinfo{year}{2018}) \bibinfo{pages}{1--8}. \URLprefix
  \url{https://www.nature.com/articles/s41467-018-07004-4}.
  \DOIprefix\doi{10.1038/s41467-018-07004-4}.
%Type = Article
\bibitem[{Palm et~al.(2024)Palm, Ding, Huxter, Taniguchi, Watanabe, and
  Degen}]{whirlpool_2024_science}
\bibinfo{author}{M.~L. Palm}, \bibinfo{author}{C.~Ding}, \bibinfo{author}{W.~S.
  Huxter}, \bibinfo{author}{T.~Taniguchi}, \bibinfo{author}{K.~Watanabe},
  \bibinfo{author}{C.~L. Degen},
\newblock \bibinfo{title}{Observation of current whirlpools in graphene at room
  temperature},
\newblock \bibinfo{journal}{Science} \bibinfo{volume}{384}
  (\bibinfo{year}{2024}) \bibinfo{pages}{465--469}. \URLprefix
  \url{https://www.science.org/doi/abs/10.1126/science.adj2167}.
  \DOIprefix\doi{10.1126/science.adj2167}.
%Type = Article
\bibitem[{Bandurin et~al.(2016)Bandurin, Torre, Kumar, Shalom, Tomadin,
  Principi, Auton, Khestanova, Novoselov, Grigorieva, Ponomarenko, Geim, and
  Polini}]{bandurin2016b}
\bibinfo{author}{D.~A. Bandurin}, \bibinfo{author}{I.~Torre},
  \bibinfo{author}{R.~K. Kumar}, \bibinfo{author}{M.~B. Shalom},
  \bibinfo{author}{A.~Tomadin}, \bibinfo{author}{A.~Principi},
  \bibinfo{author}{G.~H. Auton}, \bibinfo{author}{E.~Khestanova},
  \bibinfo{author}{K.~S. Novoselov}, \bibinfo{author}{I.~V. Grigorieva},
  \bibinfo{author}{L.~A. Ponomarenko}, \bibinfo{author}{A.~K. Geim},
  \bibinfo{author}{M.~Polini},
\newblock \bibinfo{title}{Negative local resistance caused by viscous electron
  backflow in graphene},
\newblock \bibinfo{journal}{Science} \bibinfo{volume}{351}
  (\bibinfo{year}{2016}) \bibinfo{pages}{1055--1058}. \URLprefix
  \url{http://science.sciencemag.org/content/351/6277/1055}.
  \DOIprefix\doi{10.1126/science.aad0201}.
%Type = Article
\bibitem[{Levin et~al.(2018)Levin, Gusev, Levinson, Kvon, and
  Bakarov}]{PhysRevB.97.245308}
\bibinfo{author}{A.~D. Levin}, \bibinfo{author}{G.~M. Gusev},
  \bibinfo{author}{E.~V. Levinson}, \bibinfo{author}{Z.~D. Kvon},
  \bibinfo{author}{A.~K. Bakarov},
\newblock \bibinfo{title}{Vorticity-induced negative nonlocal resistance in a
  viscous two-dimensional electron system},
\newblock \bibinfo{journal}{Phys. Rev. B} \bibinfo{volume}{97}
  (\bibinfo{year}{2018}) \bibinfo{pages}{245308}. \URLprefix
  \url{https://link.aps.org/doi/10.1103/PhysRevB.97.245308}.
  \DOIprefix\doi{10.1103/PhysRevB.97.245308}.
%Type = Article
\bibitem[{Moll et~al.(2016)Moll, Kushwaha, Nandi, Schmidt, and
  Mackenzie}]{science_2016_hydrodynamicsPdCo}
\bibinfo{author}{P.~J.~W. Moll}, \bibinfo{author}{P.~Kushwaha},
  \bibinfo{author}{N.~Nandi}, \bibinfo{author}{B.~Schmidt},
  \bibinfo{author}{A.~P. Mackenzie},
\newblock \bibinfo{title}{Evidence for hydrodynamic electron flow in
  pdcoo<sub>2</sub>},
\newblock \bibinfo{journal}{Science} \bibinfo{volume}{351}
  (\bibinfo{year}{2016}) \bibinfo{pages}{1061--1064}. \URLprefix
  \url{https://www.science.org/doi/abs/10.1126/science.aac8385}.
  \DOIprefix\doi{10.1126/science.aac8385}.
%Type = Article
\bibitem[{Aharon-Steinberg et~al.(2022)Aharon-Steinberg, V{\"o}lkl, Kaplan,
  Pariari, Roy, Holder, Wolf, Meltzer, Myasoedov, Huber
  et~al.}]{aharon2022direct}
\bibinfo{author}{A.~Aharon-Steinberg}, \bibinfo{author}{T.~V{\"o}lkl},
  \bibinfo{author}{A.~Kaplan}, \bibinfo{author}{A.~K. Pariari},
  \bibinfo{author}{I.~Roy}, \bibinfo{author}{T.~Holder},
  \bibinfo{author}{Y.~Wolf}, \bibinfo{author}{A.~Y. Meltzer},
  \bibinfo{author}{Y.~Myasoedov}, \bibinfo{author}{M.~E. Huber}, et~al.,
\newblock \bibinfo{title}{Direct observation of vortices in an electron fluid},
\newblock \bibinfo{journal}{Nature} \bibinfo{volume}{607}
  (\bibinfo{year}{2022}) \bibinfo{pages}{74--80}. \URLprefix
  \url{https://www.nature.com/articles/s41586-022-04794-y}.
  \DOIprefix\doi{https://doi.org/10.1038/s41586-022-04794-y}.
%Type = Article
\bibitem[{Mayzel et~al.(2019)Mayzel, Steinberg, and
  Varshney}]{stokes_hydrodynamic_2019}
\bibinfo{author}{J.~Mayzel}, \bibinfo{author}{V.~Steinberg},
  \bibinfo{author}{A.~Varshney},
\newblock \bibinfo{title}{Stokes flow analogous to viscous electron current in
  graphene},
\newblock \bibinfo{journal}{Nat. Commun.} \bibinfo{volume}{10}
  (\bibinfo{year}{2019}) \bibinfo{pages}{937}. \URLprefix
  \url{https://www.nature.com/articles/s41467-019-08916-5#Sec10}.
  \DOIprefix\doi{10.1038/s41467-019-08916-5}.
%Type = Article
\bibitem[{Berdyugin et~al.(2019)Berdyugin, Xu, Pellegrino, Kumar, Principi,
  Torre, Shalom, Taniguchi, Watanabe, Grigorieva, Polini, Geim, and
  Bandurin}]{berdyugin2019a}
\bibinfo{author}{A.~I. Berdyugin}, \bibinfo{author}{S.~G. Xu},
  \bibinfo{author}{F.~M.~D. Pellegrino}, \bibinfo{author}{R.~K. Kumar},
  \bibinfo{author}{A.~Principi}, \bibinfo{author}{I.~Torre},
  \bibinfo{author}{M.~B. Shalom}, \bibinfo{author}{T.~Taniguchi},
  \bibinfo{author}{K.~Watanabe}, \bibinfo{author}{I.~V. Grigorieva},
  \bibinfo{author}{M.~Polini}, \bibinfo{author}{A.~K. Geim},
  \bibinfo{author}{D.~A. Bandurin},
\newblock \bibinfo{title}{Measuring {{hall}} viscosity of graphene's electron
  fluid},
\newblock \bibinfo{journal}{Science} \bibinfo{volume}{364}
  (\bibinfo{year}{2019}) \bibinfo{pages}{162--165}. \URLprefix
  \url{https://science.sciencemag.org/content/364/6436/162}.
  \DOIprefix\doi{10.1126/science.aau0685}.
%Type = Article
\bibitem[{Crossno et~al.(2016)Crossno, Shi, Wang, Liu, Harzheim, Lucas,
  Sachdev, Kim, Taniguchi, Watanabe, Ohki, and Fong}]{science_2016_breakWFlaw}
\bibinfo{author}{J.~Crossno}, \bibinfo{author}{J.~K. Shi},
  \bibinfo{author}{K.~Wang}, \bibinfo{author}{X.~Liu},
  \bibinfo{author}{A.~Harzheim}, \bibinfo{author}{A.~Lucas},
  \bibinfo{author}{S.~Sachdev}, \bibinfo{author}{P.~Kim},
  \bibinfo{author}{T.~Taniguchi}, \bibinfo{author}{K.~Watanabe},
  \bibinfo{author}{T.~A. Ohki}, \bibinfo{author}{K.~C. Fong},
\newblock \bibinfo{title}{Observation of the dirac fluid and the breakdown of
  the wiedemann-franz law in graphene},
\newblock \bibinfo{journal}{Science} \bibinfo{volume}{351}
  (\bibinfo{year}{2016}) \bibinfo{pages}{1058--1061}. \URLprefix
  \url{https://www.science.org/doi/abs/10.1126/science.aad0343}.
  \DOIprefix\doi{10.1126/science.aad0343}.
%Type = Article
\bibitem[{Krishna~Kumar et~al.(2017)Krishna~Kumar, Bandurin, Pellegrino, Cao,
  Principi, Guo, Auton, Ben~Shalom, Ponomarenko, Falkovich, Watanabe,
  Taniguchi, Grigorieva, Levitov, Polini, and Geim}]{krishnakumar2017}
\bibinfo{author}{R.~Krishna~Kumar}, \bibinfo{author}{D.~A. Bandurin},
  \bibinfo{author}{F.~M.~D. Pellegrino}, \bibinfo{author}{Y.~Cao},
  \bibinfo{author}{A.~Principi}, \bibinfo{author}{H.~Guo},
  \bibinfo{author}{G.~H. Auton}, \bibinfo{author}{M.~Ben~Shalom},
  \bibinfo{author}{L.~A. Ponomarenko}, \bibinfo{author}{G.~Falkovich},
  \bibinfo{author}{K.~Watanabe}, \bibinfo{author}{T.~Taniguchi},
  \bibinfo{author}{I.~V. Grigorieva}, \bibinfo{author}{L.~S. Levitov},
  \bibinfo{author}{M.~Polini}, \bibinfo{author}{A.~K. Geim},
\newblock \bibinfo{title}{Superballistic flow of viscous electron fluid through
  graphene constrictions},
\newblock \bibinfo{journal}{Nat. Phys.} \bibinfo{volume}{13}
  (\bibinfo{year}{2017}) \bibinfo{pages}{1182--1185}. \URLprefix
  \url{https://www.nature.com/articles/nphys4240}.
  \DOIprefix\doi{10.1038/nphys4240}.
%Type = Article
\bibitem[{Estrada-{\'A}lvarez et~al.(2024)Estrada-{\'A}lvarez,
  Dom{\'\i}nguez-Adame, and D{\'\i}az}]{estrada2024alternative}
\bibinfo{author}{J.~Estrada-{\'A}lvarez},
  \bibinfo{author}{F.~Dom{\'\i}nguez-Adame}, \bibinfo{author}{E.~D{\'\i}az},
\newblock \bibinfo{title}{Alternative routes to electron hydrodynamics},
\newblock \bibinfo{journal}{Commun. Phys.} \bibinfo{volume}{7}
  (\bibinfo{year}{2024}) \bibinfo{pages}{138}.
  \DOIprefix\doi{https://doi.org/10.1038/s42005-024-01632-7}.
%Type = Article
\bibitem[{Scaffidi et~al.(2017)Scaffidi, Nandi, Schmidt, Mackenzie, and
  Moore}]{scaffidi2017}
\bibinfo{author}{T.~Scaffidi}, \bibinfo{author}{N.~Nandi},
  \bibinfo{author}{B.~Schmidt}, \bibinfo{author}{A.~P. Mackenzie},
  \bibinfo{author}{J.~E. Moore},
\newblock \bibinfo{title}{Hydrodynamic electron flow and hall viscosity},
\newblock \bibinfo{journal}{Phys. Rev. Lett.} \bibinfo{volume}{118}
  (\bibinfo{year}{2017}) \bibinfo{pages}{226601}. \URLprefix
  \url{https://link.aps.org/doi/10.1103/PhysRevLett.118.226601}.
  \DOIprefix\doi{10.1103/PhysRevLett.118.226601}.
%Type = Article
\bibitem[{Levitov and Falkovich(2016)}]{levitov_electron_2016}
\bibinfo{author}{L.~Levitov}, \bibinfo{author}{G.~Falkovich},
\newblock \bibinfo{title}{Electron viscosity, current vortices and negative
  nonlocal resistance in graphene},
\newblock \bibinfo{journal}{Nat. Phys.} \bibinfo{volume}{12}
  (\bibinfo{year}{2016}) \bibinfo{pages}{672--676}. \URLprefix
  \url{https://www.nature.com/articles/nphys3667}.
  \DOIprefix\doi{10.1038/nphys3667}.
%Type = Article
\bibitem[{Lucas(2017)}]{PhysRevB.95.115425}
\bibinfo{author}{A.~Lucas},
\newblock \bibinfo{title}{Stokes paradox in electronic fermi liquids},
\newblock \bibinfo{journal}{Phys. Rev. B} \bibinfo{volume}{95}
  (\bibinfo{year}{2017}) \bibinfo{pages}{115425}. \URLprefix
  \url{https://link.aps.org/doi/10.1103/PhysRevB.95.115425}.
  \DOIprefix\doi{10.1103/PhysRevB.95.115425}.
%Type = Article
\bibitem[{Vijayakrishnan et~al.(2025)Vijayakrishnan, Berkson-Korenberg,
  Mainville, Engel, Lilly, West, Pfeiffer, and
  Gervais}]{PhysRevResearch.7.L022029}
\bibinfo{author}{S.~Vijayakrishnan}, \bibinfo{author}{Z.~Berkson-Korenberg},
  \bibinfo{author}{J.~Mainville}, \bibinfo{author}{L.~W. Engel},
  \bibinfo{author}{M.~P. Lilly}, \bibinfo{author}{K.~W. West},
  \bibinfo{author}{L.~N. Pfeiffer}, \bibinfo{author}{G.~Gervais},
\newblock \bibinfo{title}{Two-dimensional hydrodynamic viscous electron flow in
  annular corbino rings},
\newblock \bibinfo{journal}{Phys. Rev. Res.} \bibinfo{volume}{7}
  (\bibinfo{year}{2025}) \bibinfo{pages}{L022029}. \URLprefix
  \url{https://link.aps.org/doi/10.1103/PhysRevResearch.7.L022029}.
  \DOIprefix\doi{10.1103/PhysRevResearch.7.L022029}.
%Type = Article
\bibitem[{Estrada-\'Alvarez et~al.(2025)Estrada-\'Alvarez,
  Dom\'{\i}nguez-Adame, and D\'{\i}az}]{PhysRevResearch.7.013087}
\bibinfo{author}{J.~Estrada-\'Alvarez},
  \bibinfo{author}{F.~Dom\'{\i}nguez-Adame}, \bibinfo{author}{E.~D\'{\i}az},
\newblock \bibinfo{title}{Anisotropic signatures of electron hydrodynamics},
\newblock \bibinfo{journal}{Phys. Rev. Res.} \bibinfo{volume}{7}
  (\bibinfo{year}{2025}) \bibinfo{pages}{013087}. \URLprefix
  \url{https://link.aps.org/doi/10.1103/PhysRevResearch.7.013087}.
  \DOIprefix\doi{10.1103/PhysRevResearch.7.013087}.
%Type = Article
\bibitem[{Talanov et~al.(2024)Talanov, Waissman, Hui, Skinner, Watanabe,
  Taniguchi, and Kim}]{corbino2024}
\bibinfo{author}{A.~Talanov}, \bibinfo{author}{J.~Waissman},
  \bibinfo{author}{A.~Hui}, \bibinfo{author}{B.~Skinner},
  \bibinfo{author}{K.~Watanabe}, \bibinfo{author}{T.~Taniguchi},
  \bibinfo{author}{P.~Kim},
\newblock \bibinfo{title}{Observation of electronic viscous dissipation in
  graphene magneto-thermal transport},
\newblock \bibinfo{journal}{arXiv.2406.13799}  (\bibinfo{year}{2024}).
  \URLprefix \url{https://doi.org/10.48550/arXiv.2406.13799}.
  \DOIprefix\doi{10.48550/arXiv.2406.13799}.
%Type = Article
\bibitem[{Gall et~al.(2023)Gall, Narozhny, and Gornyi}]{PhysRevB.107.235401}
\bibinfo{author}{V.~Gall}, \bibinfo{author}{B.~N. Narozhny},
  \bibinfo{author}{I.~V. Gornyi},
\newblock \bibinfo{title}{Corbino magnetoresistance in neutral graphene},
\newblock \bibinfo{journal}{Phys. Rev. B} \bibinfo{volume}{107}
  (\bibinfo{year}{2023}) \bibinfo{pages}{235401}. \URLprefix
  \url{https://link.aps.org/doi/10.1103/PhysRevB.107.235401}.
  \DOIprefix\doi{10.1103/PhysRevB.107.235401}.
%Type = Article
\bibitem[{Varnavides et~al.(2020)Varnavides, Jermyn, Anikeeva, Felser, and
  Narang}]{anisotropic_hydrodynamics_2020}
\bibinfo{author}{G.~Varnavides}, \bibinfo{author}{A.~S. Jermyn},
  \bibinfo{author}{P.~Anikeeva}, \bibinfo{author}{C.~Felser},
  \bibinfo{author}{P.~Narang},
\newblock \bibinfo{title}{Electron hydrodynamics in anisotropic materials},
\newblock \bibinfo{journal}{Nat. Commun.} \bibinfo{volume}{11}
  (\bibinfo{year}{2020}) \bibinfo{pages}{4710}. \URLprefix
  \url{https://www.nature.com/articles/s41467-020-18553-y}.
  \DOIprefix\doi{10.1038/s41467-020-18553-y}.
%Type = Article
\bibitem[{Vijayakrishnan et~al.(2023)Vijayakrishnan, Poitevin, Yu,
  Berkson-Korenberg, Petrescu, Lilly, Szkopek, Agarwal, West, Pfeiffer, and
  Gervais}]{anomalous_2023_corbino}
\bibinfo{author}{S.~Vijayakrishnan}, \bibinfo{author}{F.~Poitevin},
  \bibinfo{author}{O.~Yu}, \bibinfo{author}{Z.~Berkson-Korenberg},
  \bibinfo{author}{M.~Petrescu}, \bibinfo{author}{M.~P. Lilly},
  \bibinfo{author}{T.~Szkopek}, \bibinfo{author}{K.~Agarwal},
  \bibinfo{author}{K.~W. West}, \bibinfo{author}{L.~N. Pfeiffer},
  \bibinfo{author}{G.~Gervais},
\newblock \bibinfo{title}{Anomalous electronic transport in high-mobility
  corbino rings},
\newblock \bibinfo{journal}{Nat. Commun.} \bibinfo{volume}{14}
  (\bibinfo{year}{2023}) \bibinfo{pages}{3906}. \URLprefix
  \url{https://www.nature.com/articles/s41467-023-39526-x}.
  \DOIprefix\doi{10.1038/s41467-023-39526-x}.
%Type = Article
\bibitem[{Falkovich and Levitov(2017)}]{falkovich_linking_2017}
\bibinfo{author}{G.~Falkovich}, \bibinfo{author}{L.~Levitov},
\newblock \bibinfo{title}{Linking spatial distributions of potential and
  current in viscous electronics},
\newblock \bibinfo{journal}{Phys. Rev. Lett.} \bibinfo{volume}{119}
  (\bibinfo{year}{2017}) \bibinfo{pages}{066601}. \URLprefix
  \url{https://link.aps.org/doi/10.1103/PhysRevLett.119.066601}.
  \DOIprefix\doi{10.1103/PhysRevLett.119.066601}.
%Type = Article
\bibitem[{Torre et~al.(2015)Torre, Tomadin, Geim, and Polini}]{torre2015}
\bibinfo{author}{I.~Torre}, \bibinfo{author}{A.~Tomadin},
  \bibinfo{author}{A.~K. Geim}, \bibinfo{author}{M.~Polini},
\newblock \bibinfo{title}{Nonlocal transport and the hydrodynamic shear
  viscosity in graphene},
\newblock \bibinfo{journal}{Phys. Rev. B} \bibinfo{volume}{92}
  (\bibinfo{year}{2015}) \bibinfo{pages}{165433}. \URLprefix
  \url{https://link.aps.org/doi/10.1103/PhysRevB.92.165433}.
  \DOIprefix\doi{10.1103/PhysRevB.92.165433}.
%Type = Article
\bibitem[{Pellegrino et~al.(2016)Pellegrino, Torre, Geim, and
  Polini}]{PhysRevB.94.155414}
\bibinfo{author}{F.~M.~D. Pellegrino}, \bibinfo{author}{I.~Torre},
  \bibinfo{author}{A.~K. Geim}, \bibinfo{author}{M.~Polini},
\newblock \bibinfo{title}{Electron hydrodynamics dilemma: Whirlpools or no
  whirlpools},
\newblock \bibinfo{journal}{Phys. Rev. B} \bibinfo{volume}{94}
  (\bibinfo{year}{2016}) \bibinfo{pages}{155414}. \URLprefix
  \url{https://link.aps.org/doi/10.1103/PhysRevB.94.155414}.
  \DOIprefix\doi{10.1103/PhysRevB.94.155414}.
%Type = Article
\bibitem[{Keser et~al.(2021)Keser, Wang, Klochan, Ho, Tkachenko, Tkachenko,
  Culcer, Adam, Farrer, Ritchie, Sushkov, and Hamilton}]{PhysRevX.11.031030}
\bibinfo{author}{A.~i. e. i. f.~C. Keser}, \bibinfo{author}{D.~Q. Wang},
  \bibinfo{author}{O.~Klochan}, \bibinfo{author}{D.~Y.~H. Ho},
  \bibinfo{author}{O.~A. Tkachenko}, \bibinfo{author}{V.~A. Tkachenko},
  \bibinfo{author}{D.~Culcer}, \bibinfo{author}{S.~Adam},
  \bibinfo{author}{I.~Farrer}, \bibinfo{author}{D.~A. Ritchie},
  \bibinfo{author}{O.~P. Sushkov}, \bibinfo{author}{A.~R. Hamilton},
\newblock \bibinfo{title}{Geometric control of universal hydrodynamic flow in a
  two-dimensional electron fluid},
\newblock \bibinfo{journal}{Phys. Rev. X} \bibinfo{volume}{11}
  (\bibinfo{year}{2021}) \bibinfo{pages}{031030}. \URLprefix
  \url{https://link.aps.org/doi/10.1103/PhysRevX.11.031030}.
  \DOIprefix\doi{10.1103/PhysRevX.11.031030}.
%Type = Article
\bibitem[{Li et~al.(2022)Li, Levchenko, and Andreev}]{PhysRevB.105.125302}
\bibinfo{author}{S.~Li}, \bibinfo{author}{A.~Levchenko}, \bibinfo{author}{A.~V.
  Andreev},
\newblock \bibinfo{title}{Hydrodynamic thermoelectric transport in corbino
  geometry},
\newblock \bibinfo{journal}{Phys. Rev. B} \bibinfo{volume}{105}
  (\bibinfo{year}{2022}) \bibinfo{pages}{125302}. \URLprefix
  \url{https://link.aps.org/doi/10.1103/PhysRevB.105.125302}.
  \DOIprefix\doi{10.1103/PhysRevB.105.125302}.
%Type = Article
\bibitem[{Gall et~al.(2023)Gall, Narozhny, and Gornyi}]{PhysRevB.107.045413}
\bibinfo{author}{V.~Gall}, \bibinfo{author}{B.~N. Narozhny},
  \bibinfo{author}{I.~V. Gornyi},
\newblock \bibinfo{title}{Electronic viscosity and energy relaxation in neutral
  graphene},
\newblock \bibinfo{journal}{Phys. Rev. B} \bibinfo{volume}{107}
  (\bibinfo{year}{2023}) \bibinfo{pages}{045413}. \URLprefix
  \url{https://link.aps.org/doi/10.1103/PhysRevB.107.045413}.
  \DOIprefix\doi{10.1103/PhysRevB.107.045413}.
%Type = Article
\bibitem[{Yao et~al.(2026)Yao, Yang, Yang, and Qian}]{yao2026imaging}
\bibinfo{author}{Y.~Yao}, \bibinfo{author}{H.~Yang}, \bibinfo{author}{R.~Yang},
  \bibinfo{author}{X.~Qian},
\newblock \bibinfo{title}{Imaging thermal properties of thermal interface
  materials using frequency-domain thermoreflectance microscopy},
\newblock \bibinfo{journal}{Thermo-X} \bibinfo{volume}{2}
  (\bibinfo{year}{2026}) \bibinfo{pages}{202612}. \URLprefix
  \url{https://doi.org/10.70401/tx.2026.0020}.
%Type = Article
\bibitem[{Zhang et~al.(2026)Zhang, Fan, Ma, and Zhang}]{zhangx_2026_thermoX}
\bibinfo{author}{Y.~Zhang}, \bibinfo{author}{A.~Fan}, \bibinfo{author}{W.~Ma},
  \bibinfo{author}{X.~Zhang},
\newblock \bibinfo{title}{Interfacial heat transport in two-dimensional
  heterostructures: From formation to functionality},
\newblock \bibinfo{journal}{Thermo-X} \bibinfo{volume}{2}
  (\bibinfo{year}{2026}) \bibinfo{pages}{202620}. \URLprefix
  \url{https://doi.org/10.70401/tx.2026.0021}.
%Type = Article
\bibitem[{Narozhny(2019)}]{NAROZHNY2019167979}
\bibinfo{author}{B.~N. Narozhny},
\newblock \bibinfo{title}{Electronic hydrodynamics in graphene},
\newblock \bibinfo{journal}{Annals of Physics} \bibinfo{volume}{411}
  (\bibinfo{year}{2019}) \bibinfo{pages}{167979}. \URLprefix
  \url{https://www.sciencedirect.com/science/article/pii/S0003491619302349}.
  \DOIprefix\doi{https://doi.org/10.1016/j.aop.2019.167979}.
%Type = Article
\bibitem[{Gooth et~al.(2018)Gooth, Menges, Kumar, S{\"u}$\beta$, Shekhar, Sun,
  Drechsler, Zierold, Felser, and Gotsmann}]{gooth2018thermal}
\bibinfo{author}{J.~Gooth}, \bibinfo{author}{F.~Menges},
  \bibinfo{author}{N.~Kumar}, \bibinfo{author}{V.~S{\"u}$\beta$},
  \bibinfo{author}{C.~Shekhar}, \bibinfo{author}{Y.~Sun},
  \bibinfo{author}{U.~Drechsler}, \bibinfo{author}{R.~Zierold},
  \bibinfo{author}{C.~Felser}, \bibinfo{author}{B.~Gotsmann},
\newblock \bibinfo{title}{Thermal and electrical signatures of a hydrodynamic
  electron fluid in tungsten diphosphide},
\newblock \bibinfo{journal}{Nat. Commun.} \bibinfo{volume}{9}
  (\bibinfo{year}{2018}) \bibinfo{pages}{4093}. \URLprefix
  \url{https://www.nature.com/articles/s41467-018-06688-y}.
  \DOIprefix\doi{https://doi.org/10.1038/s41467-018-06688-y}.
%Type = Article
\bibitem[{Li and Levchenko(2022)}]{PhysRevB.105.L241405}
\bibinfo{author}{S.~Li}, \bibinfo{author}{A.~Levchenko},
\newblock \bibinfo{title}{Nonlocal thermoelectric resistance in vortical
  viscous transport},
\newblock \bibinfo{journal}{Phys. Rev. B} \bibinfo{volume}{105}
  (\bibinfo{year}{2022}) \bibinfo{pages}{L241405}. \URLprefix
  \url{https://link.aps.org/doi/10.1103/PhysRevB.105.L241405}.
  \DOIprefix\doi{10.1103/PhysRevB.105.L241405}.
%Type = Article
\bibitem[{Chen(2021)}]{chen_non-fourier_2021}
\bibinfo{author}{G.~Chen},
\newblock \bibinfo{title}{Non-{Fourier} phonon heat conduction at the
  microscale and nanoscale},
\newblock \bibinfo{journal}{Nat. Rev. Phys.} \bibinfo{volume}{3}
  (\bibinfo{year}{2021}) \bibinfo{pages}{555--569}. \URLprefix
  \url{https://www.nature.com/articles/s42254-021-00334-1}.
  \DOIprefix\doi{10.1038/s42254-021-00334-1}.
%Type = Article
\bibitem[{Matsuura et~al.(2026)Matsuura, Takahashi, and
  Katase}]{APR2026_phonondrag}
\bibinfo{author}{H.~Matsuura}, \bibinfo{author}{H.~Takahashi},
  \bibinfo{author}{T.~Katase},
\newblock \bibinfo{title}{History and perspective of phonon–drag effect
  advancing thermoelectrics},
\newblock \bibinfo{journal}{Applied Physics Reviews} \bibinfo{volume}{13}
  (\bibinfo{year}{2026}) \bibinfo{pages}{011321}. \URLprefix
  \url{https://doi.org/10.1063/5.0312585}. \DOIprefix\doi{10.1063/5.0312585}.
%Type = Book
\bibitem[{Kaviany(2008)}]{kaviany_2008}
\bibinfo{author}{M.~Kaviany}, \bibinfo{title}{Heat transfer physics},
  \bibinfo{publisher}{Cambridge University Press}, \bibinfo{year}{2008}.
  \URLprefix \url{https://doi.org/10.1017/CBO9780511754586}.
  \DOIprefix\doi{10.1017/CBO9780511754586}.
%Type = Article
\bibitem[{Zhang et~al.(2017)Zhang, Guo, and Chen}]{Chuang17gray}
\bibinfo{author}{C.~Zhang}, \bibinfo{author}{Z.~Guo},
  \bibinfo{author}{S.~Chen},
\newblock \bibinfo{title}{Unified implicit kinetic scheme for steady multiscale
  heat transfer based on the phonon {B}oltzmann transport equation},
\newblock \bibinfo{journal}{Phys. Rev. E} \bibinfo{volume}{96}
  (\bibinfo{year}{2017}) \bibinfo{pages}{063311}. \URLprefix
  \url{https://link.aps.org/doi/10.1103/PhysRevE.96.063311}.
  \DOIprefix\doi{10.1103/PhysRevE.96.063311}.
%Type = Article
\bibitem[{Yoon and Jameson(1988)}]{YoonS88LUSGS}
\bibinfo{author}{S.~Yoon}, \bibinfo{author}{A.~Jameson},
\newblock \bibinfo{title}{Lower-upper symmetric-gauss-seidel method for the
  euler and navier-stokes equations},
\newblock \bibinfo{journal}{AIAA Journal} \bibinfo{volume}{26}
  (\bibinfo{year}{1988}) \bibinfo{pages}{1025--1026}. \URLprefix
  \url{http://arc.aiaa.org/doi/abs/10.2514/3.10007}.
  \DOIprefix\doi{10.2514/3.10007}.
%Type = Article
\bibitem[{Gupta et~al.(2021)Gupta, Heremans, Kataria, Chandra, Fallahi,
  Gardner, and Manfra}]{PhysRevLett.126.076803}
\bibinfo{author}{A.~Gupta}, \bibinfo{author}{J.~J. Heremans},
  \bibinfo{author}{G.~Kataria}, \bibinfo{author}{M.~Chandra},
  \bibinfo{author}{S.~Fallahi}, \bibinfo{author}{G.~C. Gardner},
  \bibinfo{author}{M.~J. Manfra},
\newblock \bibinfo{title}{Hydrodynamic and ballistic transport over large
  length scales in $\mathrm{GaAs}/\mathrm{AlGaAs}$},
\newblock \bibinfo{journal}{Phys. Rev. Lett.} \bibinfo{volume}{126}
  (\bibinfo{year}{2021}) \bibinfo{pages}{076803}. \URLprefix
  \url{https://link.aps.org/doi/10.1103/PhysRevLett.126.076803}.
  \DOIprefix\doi{10.1103/PhysRevLett.126.076803}.
%Type = Article
\bibitem[{Van~Leer(1977)}]{vanleer1977}
\bibinfo{author}{B.~Van~Leer},
\newblock \bibinfo{title}{Towards the ultimate conservative difference scheme.
  {{IV}}. {{A}} new approach to numerical convection},
\newblock \bibinfo{journal}{J. Comput. Phys.} \bibinfo{volume}{23}
  (\bibinfo{year}{1977}) \bibinfo{pages}{276--299}. \URLprefix
  \url{http://www.sciencedirect.com/science/article/pii/002199917790095X}.
  \DOIprefix\doi{10.1016/0021-9991(77)90095-X}.
%Type = Article
\bibitem[{Miao et~al.(2019)Miao, Guo, Ran, and Wang}]{PhysRevB.99.205433}
\bibinfo{author}{W.~Miao}, \bibinfo{author}{Y.~Guo}, \bibinfo{author}{X.~Ran},
  \bibinfo{author}{M.~Wang},
\newblock \bibinfo{title}{Deviational monte carlo scheme for thermal and
  electrical transport in metal nanostructures},
\newblock \bibinfo{journal}{Phys. Rev. B} \bibinfo{volume}{99}
  (\bibinfo{year}{2019}) \bibinfo{pages}{205433}. \URLprefix
  \url{https://link.aps.org/doi/10.1103/PhysRevB.99.205433}.
  \DOIprefix\doi{10.1103/PhysRevB.99.205433}.
%Type = Article
\bibitem[{Fuchs(1938)}]{Fuchs_1938_analytical}
\bibinfo{author}{K.~Fuchs},
\newblock \bibinfo{title}{The conductivity of thin metallic films according to
  the electron theory of metals},
\newblock \bibinfo{journal}{Mathematical Proceedings of the Cambridge
  Philosophical Society} \bibinfo{volume}{34} (\bibinfo{year}{1938})
  \bibinfo{pages}{100–108}. \DOIprefix\doi{10.1017/S0305004100019952}.
%Type = Article
\bibitem[{Sondheimer(1952)}]{Sondheimer_1952_analytical}
\bibinfo{author}{E.~Sondheimer},
\newblock \bibinfo{title}{The mean free path of electrons in metals},
\newblock \bibinfo{journal}{Advances in Physics} \bibinfo{volume}{1}
  (\bibinfo{year}{1952}) \bibinfo{pages}{1--42}. \URLprefix
  \url{https://doi.org/10.1080/00018735200101151}.
  \DOIprefix\doi{10.1080/00018735200101151}.

\end{thebibliography}

\end{document}